\documentclass[twocolumn]{aastex61}
\usepackage{amssymb,latexsym}
\usepackage{graphicx}
\usepackage{amsmath}
\usepackage{xcolor}

\usepackage{array}
\usepackage[]{natbib}

\usepackage[english]{babel}
\usepackage{blindtext}
\usepackage{lipsum}

\graphicspath{{./figures/}}

\begin{document}

\title{A multimass velocity dispersion model of 47 Tucanae indicates no evidence for an intermediate mass black hole}

\author{Christopher R. Mann}
\affil{Department of Physics and Astronomy\\
         \ University of British Columbia\\
         \ Vancouver, BC V6T 1Z1\\
         \ Canada} 
       
\author{Harvey Richer}
\affil{Department of Physics and Astronomy\\
         \ University of British Columbia\\
         \ Vancouver, BC V6T 1Z1\\
         \ Canada} 
	       
\author{Jeremy Heyl}
\affil{Department of Physics and Astronomy\\
         \ University of British Columbia\\
         \ Vancouver, BC V6T 1Z1\\
         \ Canada} 
         
\author{Jay Anderson}
\affil{Space Telescope Science Institute\\
		\ 3700 San Martin Drive\\
		\ Baltimore, MD 21218\\
		\ USA }     

\author{Jason Kalirai}
\affil{Space Telescope Science Institute\\
		\ 3700 San Martin Drive\\
		\ Baltimore, MD 21218\\
		\ USA }      

\author{Ilaria Caiazzo}
\affil{Department of Physics and Astronomy\\
         \ University of British Columbia\\
         \ Vancouver, BC V6T 1Z1\\
         \ Canada}    
         
\author{Swantje D. M\"{o}hle}
\affil{Department of Physics and Astronomy\\
         \ University of British Columbia\\
         \ Vancouver, BC V6T 1Z1\\
         \ Canada} 
         
\author{Alan Knee}
\affil{Department of Physics and Astronomy\\
         \ University of British Columbia\\
         \ Vancouver, BC V6T 1Z1\\
         \ Canada} 

\author{Holger Baumgardt}
\affil{School of Mathematics and Physics\\
		\ University of Queensland\\
		\ St. Lucia, QLD 4068\\
		\ Australia}

\begin{abstract}

In this paper, we analyze stellar proper motions in the core of the globular cluster 47 Tucanae to explore the possibility of an intermediate-mass black hole (IMBH) influence on the stellar dynamics.
Our use of short-wavelength photometry affords us an exceedingly clear view of stellar motions into the very center of the
crowded core, yielding proper motions for $>$50,000 stars in the central 2$^{\prime}$.  We model the velocity dispersion profile of the cluster using an isotropic Jeans model. The density distribution is taken as a central IMBH point mass added to a combination of King templates.  
We individually model the general low-mass cluster objects (main-sequence/giant stars), as well as the concentrated populations of heavy binary systems and dark stellar remnants.
Using unbinned likelihood model fitting, we find that the inclusion of 
the concentrated populations in our model plays a crucial role in fitting for an IMBH mass.
The concentrated binaries and stellar-mass black holes (BHs) produce a sufficient velocity dispersion signal in the core so as to make an IMBH unnecessary to fit the observations.
We additionally determine that a stellar-mass BH retention fraction of $\gtrsim 8.5\%$
becomes incompatible with our observed velocities in the core.

\end{abstract}


\section{Introduction}\label{S:intro}

There are two classes of black holes (BHs) for which we have strong observational evidence.  Stellar-mass BHs (sBHs) are of the order of a few solar masses, while supermassive BHs in galactic cores are found to be in the range of $10^6-10^9\ M_\odot$ \citep[for example, see][for reviews]{Ferrarese_Ford_2005,Graham_2016}.  Bridging the gap between these two extremes is the regime of the intermediate-mass BH (IMBH).  Though IMBHs are required by most theories of supermassive BH evolution \citep{Mezcua_2017}, they have proven very difficult to confirm observationally.
There are many proposed formation mechanisms
\citep[e.g.][]{QuinlanShapiro1990,SigurdssonHernquist1993,Fryer1999,MillerHamilton2002,GurkanEtal2004,PortegiesZwartEtal2004,FreitagEtal2006,Giersz_etal_2015}, but most require a dense stellar environment of the type that is typically found only in the cores of globular clusters.  As such, globular clusters tend to be the primary targets for IMBH searches. 

One of the many open questions surrounding IMBHs is whether or not they follow the same empirical bulge-BH relations observed for supermassive BHs in galaxies.  Correlations are present between the mass of the central supermassive BH and total bulge mass, total bulge luminosity, and central bulge velocity dispersion \citep[][and references therein]{Graham_2016}.
Such correlations suggest commensurate evolution and/or feedback between the BH and its surroundings.  
IMBHs may follow the same or similar relations with their host globular clusters \citep{LutzgendorfEtal2013b}.
Discrepancies between the bulge-BH relations of supermassive BHs versus IMBHs may shed light on how IMBHs themselves form, and how they are involved in the growth of supermassive BHs \citep{Mezcua_2017}.

There are several methods available to detect IMBHs.  Systems actively accreting gas may show corresponding X-ray and radio emission. 
The detected flux combined with accretion models places an upper limit on the mass of the accreting BH.  This method is frequently used to constrain stellar-mass BHs in observed X-ray binaries as well as for galactic supermassive BHs.  The technique can be applied to IMBHs using the BH accretion fundamental plane, an empirical correlation among a BH's radio flux, X-ray flux, and mass \citep{Merloni_etal_2003}.  However, this relation has not yet been explicitly confirmed to continue into the IMBH mass regime.  
Our ability to look for radio signatures of IMBHs in globular clusters will be dramatically expanded when the Next Generation Very Large Array comes online in roughly a decade, as described in a proposal by \citet{Wrobel_etal_2018}.  This proposal outlines a plan to measure the radio fluxes of many hundreds of globular clusters out to a distance of 25 Mpc.
Reaching large numbers of globular clusters is important as it may be the case that only a small fraction of clusters are likely to keep an IMBH even if they are commonly born with one \citep{Fragione_etal_2018}.

The kinematic study of stellar clusters is a common approach for central IMBH detection.
One way to probe the gravitational environment is to examine pulsar accelerations using measurements of their various spin derivatives \citep{KiziltanEtal2017,Gieles_etal_2018}. There are, however, confounding factors such as the intrinsic spindown of the pulsars.
Of particular interest to the study at hand are the results for 47 Tuc by \citet{KiziltanEtal2017}.
They determined accelerations of millisecond pulsars in 47 Tuc and compared them against $N$-body simulations with different masses of central IMBHs.  
Their analysis suggested that an IMBH of  mass 
$2300^{+1500}_{-850}\ M_\odot$ 
($0.30\% ^{+0.20\%}_{-0.12\%}$ 
of their total cluster mass) may be present in order to explain their observed pulsar accelerations.
We would like to note some deficiencies with certain aspects of their method. The simulations used for comparison with their pulsar data contained no significant mass in binaries. They included no primordial binary systems, leaving only those systems that formed dynamically.  
In a cluster with a short central relaxation time such as 47 Tuc, the binaries and other massive objects will be centrally concentrated.
As we demonstrate below, the presence of a concentrated and substantial mass distribution can mimic the dynamical effect of an IMBH.  Including binaries is crucial for a cluster like 47 Tuc, the binary mass fraction of which exceeds a few percent of the entire cluster mass \citep{MiloneEtal2012}. \citet{KiziltanEtal2017} also underestimated the neutron star (NS) masses in their simulation, which they took as the canonical $1.4 M_\odot$.  
Most millisecond pulsars in 47 Tuc are part of binary systems 
\citep{Pan_etal_2016,Ridolfi_etal_2016,Freire_etal_2017}
and have total dynamical masses $\gtrsim 2\ M_\odot$.
This mass difference will affect their distribution in the cluster and must be taken into account.

Another kinematic approach to detecting an IMBH,
and the process that is employed in this paper,
is to measure and model the velocity dispersion of stars in the core of the globular cluster.  This technique is commonly used in measuring galactic supermassive BHs \citep{Walsh_etal_2013,ChatzopoulosEtal2015,AbhEtal2017,PagottoEtal2017}. 
For galactic studies, individual stars generally cannot be resolved and so gas-dynamical models and spectral line widths are used to infer velocity dispersions.
Similar techniques have produced
upper limits for IMBHs in globular clusters where individual stars can be measured in the line-of-sight velocity and/or proper motion 
(for examples, see Table~\ref{T:lit_IMBH_masses}).

However, individual stellar measurements can be easily hindered by the crowding in the core, which is highly problematic as the core is the region where an IMBH's gravitational influence will be most prevalent.
Using integrated-light spectroscopy, one can try to work around the visibility issue in the crowded globular cluster cores.  This method can be very useful, but also comes with its own limitations \citep[see, e.g.][]{DeVita_etal_2017}.
Table~\ref{T:lit_IMBH_masses} displays a selection of recent IMBH mass
estimates and limits placed on various clusters through both emission and dynamical methods. It is clear that there is a lot of disagreement, even for studies with the same target.

A central cusp in the density and/or luminosity profile of a cluster is often noted as possible
evidence for an IMBH.  However, numerical and observational studies looking for such cusps or relations between the core radius and half-light or half-mass radius have not revealed
clear-cut signatures that indicate an IMBH \citep{Baumgardt_etal_2005,Trenti_2006,Hurley_2007,NoyolaEtal2008}.
For a comprehensive review of the observational evidence for IMBHs, see \citet{Mezcua_2017}.

This paper documents our use of proper motions in the core of the globular cluster 47 Tucanae to conduct analysis on the velocity dispersion profile and investigate whether an IMBH is required to explain the observed stellar motions.  
Section~\ref{S:Data} outlines how we use ultraviolet (UV) images to overcome the visibility issues in the core when determining proper motions and how we characterize a population of dynamically heavy binary systems using an optical/near-infrared (NIR) data set.  It also outlines our process of estimating the populations of dark stellar remnants.
In Section~\ref{S:Model} we build the analytic velocity dispersion model that is fit to the observations
and discuss our methods while explaining our use of an unbinned likelihood maximization technique to fit the velocity dispersion model.
In Section~\ref{S:Results} we report our results and uncertainties, verifying the validity of our velocity dispersion model with a resampling technique.
Finally, Section~\ref{S:Summary_Discussion} provides a summary and discussion of our findings.

\begin{table}[t]  
\begin{centering}
\caption{Recent IMBH Findings/Limits} \label{T:lit_IMBH_masses}
\begin{tabular}{ p{1.9cm}   p{1.7cm}   p{1.6cm} p{0.5cm}  }
Target		& Method  	&$M_\text{IMBH}$ ($M_\odot$) & References	\\
\hline \hline
$\omega$ Cen 	& kinematic	 & $<$12,000  	& (1) \\
$\omega$ Cen	& kinematic	 & 40,000  		& (2) \\
$\omega$ Cen	& kinematic	 & lower than reference (2)$^a$ 		& (3) \\
NGC 6388		& kinematic	 & $<$2000  		& (4) \\
NGC 6388		& kinematic	 & 28,000  		& (5) \\
NGC 6388		& X-ray		 & 1500  		& (6) \\
NGC 6388		& X-ray	 	 & $<$600$^b$   $<$1200$^b$  		& (7) \\
47 Tuc			& kinematic	 & 2300  		& (8) \\
NGC 6535		& comparative simulation	 & presence$^c$ 		& (9) \\
ULX-7 (M51)		& X-ray	     & $<$1600$^b$  $<$35,000$^b$ 	& (10) \\
M15				& Radio VLBI	 & $<$500  		& (11) \\
Molec. cloud CO-0.40-0.22	& Radio, gas dynamics	 & 100,000  		& (12) \\
NGC 6624		& Pulsar timing	 & $>$7500  	& (13) \\
NGC 6624		& Pulsar timing  & \ \ 0			& (14) \\
G1 (M31)		& kinematic		& 17,000		& (15) \\
G1 (M31)		& comparative simulation	& \ \ 0				& (16) \\
NGC 6397		& kinematic		& 600				& (17) \\
\hline
\end{tabular}
\tablecomments{}{
        {\bf Notes.}
        \newline
        $^a$ Authors do not provide mass estimate, but their anisotropy analysis suggests reference (2) is too large.
        \newline
        $^b$ These estimates reflect different techniques and 
        assumptions within the referenced study.
        \newline
        $^c$ Comparison with simulations suggest the 
        presence of an IMBH, but there are no reported constraints on mass.
        \newline
        {\bf References.}
		(1) \citet{VanderMarelAnderson2010},
        (2)	\citet{NoyolaEtal2008},
        (3)	\citet{Zocchi_etal_2017},
        (4)	\citet{Lanzoni_etal_2013},
        (5)	\citet{Lutzgendorf_etal_2015},
        (6)	\citet{Cseh_etal_2010},
        (7)	\citet{Bozzo_etal_2011},
        (8)	\citet{KiziltanEtal2017},
        (9)	\citet{Askar_etal_2017},
        (10) \citet{Earnshaw_etal_2016},
        (11) \citet{Kirsten_Vlemmings_2012},
        (12) \citet{Oka_etal_2017},
        (13) \citet{Perera_etal_2017},
        (14) \citet{Gieles_etal_2018},
        (15) \citet{Gebhardt_etal_2005},
        (16) \citet{Baumgardt_etal_2003},
        (17) \citet{Kamann_etal_2016}.
        }
\end{centering}
\end{table}

\section{Data}\label{S:Data}

\subsection{Proper Motions}\label{ss:PMs}

Our proper motion data come from two epochs of \emph{Hubble Space Telescope (HST)} observations.  \emph{HST} program GO-12971 (PI: Richer) 
observed 47 Tuc over 10 orbits early in 2013, imaging the core in the F225W and F336W filters using Wide Field Camera 3.   These data were used in conjunction with those of an earlier epoch;  GO-9443 (PI: King) imaged the core in 2002 using the ACS
 High-Resolution Channel in the F475W filter.  We match the stars between the F336W and F475W images, providing a baseline of $\sim 11$ yr over which to calculate proper motions.

\begin{figure}[t]
	\centering
	\includegraphics[width=0.48\textwidth]{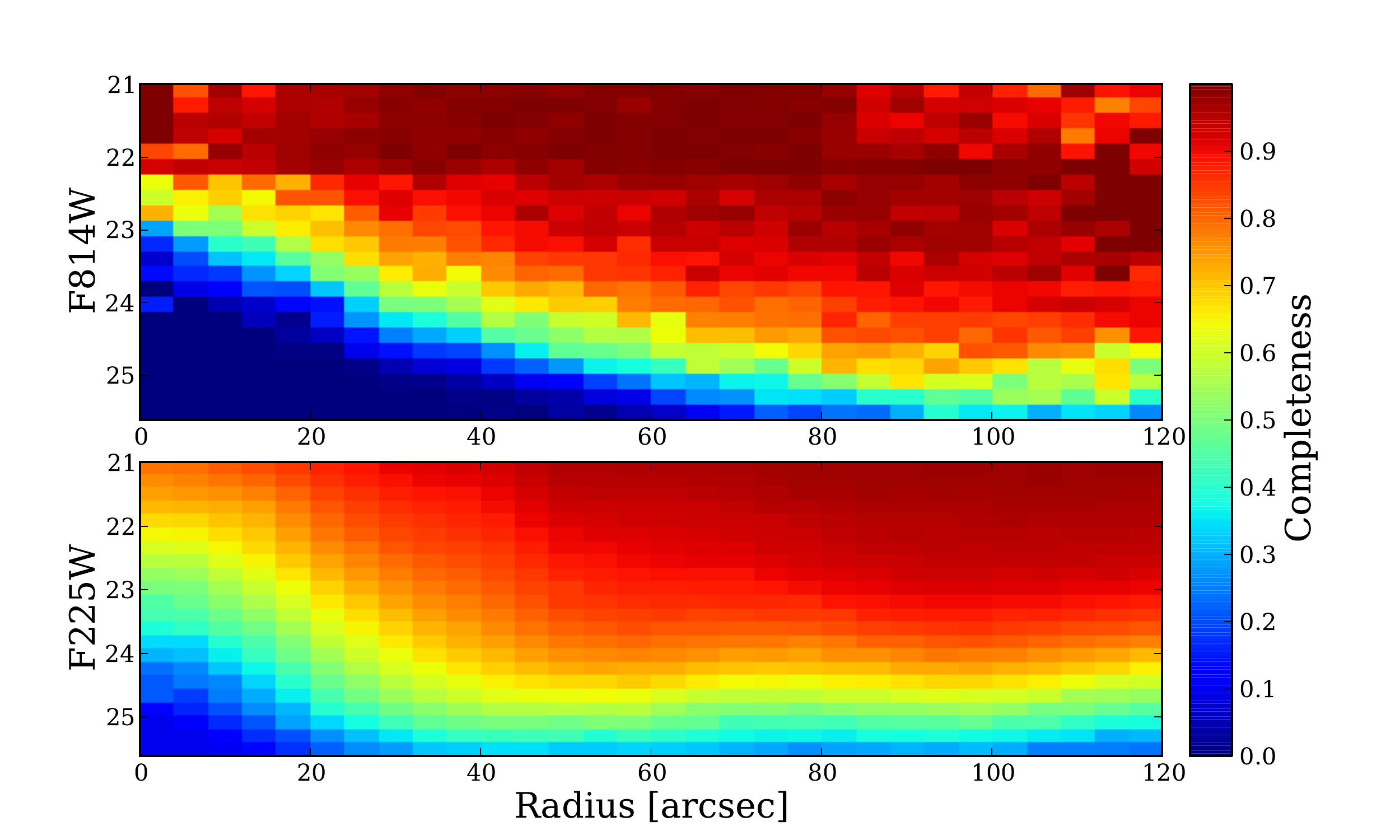}
	\caption{ 
    	Completeness results between NIR and UV photometry.
        {Top:} near-infrared data and artificial star test from \citet{Sarajedini_etal_2007}.
        {Bottom:} UV data used in this paper.  Artificial star tests described in
        \citet{GoldsburyEtal2016}.
		\label{F:Completeness}}
\end{figure}

It is worth taking a moment to note the benefits of using such blue/UV filters.  
Historically, observations in the core of 47 Tuc have tended to be at longer wavelengths.
In such filters, visibility in the central region is hampered both by crowding and by the dominating brightness of the many red giant stars located there \citep{Sarajedini_etal_2007}.
These giants saturate
a portion of the detector in any moderately deep image, bleeding outwards and
reducing the ability to accurately measure their stellar neighbors.  Our short-wavelength filters act to suppress the light from the cool red giants and allow for an unimpeded deep exposure right in the cluster's core.  The shorter wavelengths also
improve the diffraction limit of every star in the frame, reducing the degree of crowding.  
Figure~\ref{F:Completeness} displays the higher core completeness of the bluer filters.
For a detailed description of the observations and completeness correction process, see \citet{GoldsburyEtal2016}.

As a result of this extremely clear view into the core, we achieve quality proper motion measurements of over 50,000 stars, including 
$12$ within the central arcsecond,
and $100$ within the central $3^{\prime\prime}$.  Measurements this close to the cluster core have been extremely difficult in
the past, and yet these are exactly the stars that probe the region of greatest potential IMBH influence.  
A dynamical estimate \citep{Peebles_1972} sets the BH's radius of influence ($r_I$) to be where a circular orbital speed around the IMBH would equal the central cluster velocity dispersion: 
\begin{equation}
	r_{I} = \frac{GM_\text{IMBH}}{\sigma_c^2},
\end{equation}
where $\sigma_c$ is the central velocity dispersion of the cluster.
Adopting a distance to 47 Tuc of 4.69 kpc \citep{Woodley_etal_2012} and measuring the central velocity dispersion to be $\sigma_c \approx 15$ kms$^{-1}$ (value found in this study), we calculate the sphere of influence of a $2000\ M_\odot$ BH in 47 Tuc to have a radius of $r_I \approx 1.7^{\prime\prime}$.  
{
\citet{McLaughlin_etal_2006} report a line-of-sight velocity dispersion in the core somewhat lower
than that found here, though scaling our velocities to their choice of distance ($\sim 4$ kpc) largely explains the difference.
Additionally, their line-of-sight dispersion value is for the entire $R<25''$ region rather than
right in the core.
Uncertainties in distance as well as discrepancies in $\sigma_c$ measurements will affect $r_I$, but it remains clear that} the central few arcseconds are where an IMBH has its most pronounced effect, so visibility in this region is critical.


\subsection{Massive Objects}

Throughout this paper, we will be referring to different components and subpopulations of 47 Tuc with the following abbreviations in the text and quantity subscripts: IMBH, low-mass cluster objects (Cl), binaries (bin), white dwarfs (WDs), NS, and sBHs.

When looking for a velocity dispersion signature in the core, it is important to separately account for populations of dynamically massive objects.  
These could be inherently massive objects like NSs or sBHs, or they could be binary systems of lighter objects, whose combined mass causes them to behave like heavy objects in the cluster.
Relaxation effects cause more massive objects to be concentrated in the core while the less massive ones diffuse toward the outer reaches.  This segregation will prove to be important in the context of velocity dispersion and will be discussed in later sections.
For the purpose of our analysis, we make the distinction between high-mass  ($\gtrsim 1.0\ M_\odot$) and low-mass ($\lesssim 1.0\ M_\odot$) cluster objects.  
For the low-mass category, we group together the low-mass binaries, low-mass WDs, main-sequence stars, and giants as a single composite population.  
In the high-mass category, we individually model the populations of heavy binary systems, heavy WDs, NSs, and sBHs.  
These higher mass objects are more concentrated than the typical cluster object and have the potential to affect the velocity dispersion deep into the core, mimicking the effects of an IMBH
\citep{Zocchi_etal_2019}.

\subsubsection{Binary Population}\label{sS:Binaries}

\begin{figure}[t]
	\centering
	\includegraphics[width=0.48\textwidth]{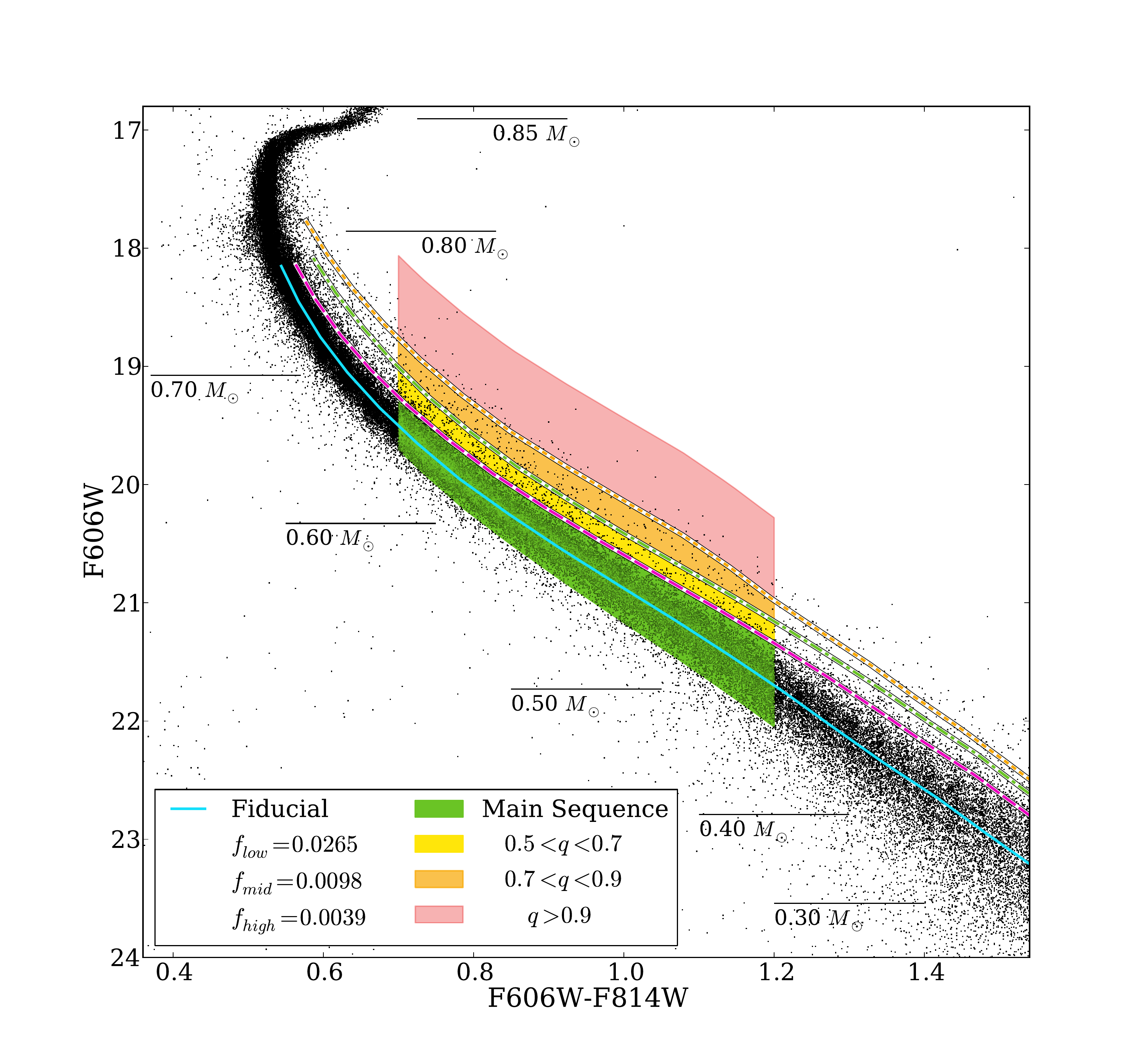}
	\caption{ 
    A color-magnitude diagram of 47 Tuc stars in visual/NIR filters. 
    We use a fiducial main-sequence line and 11 Gyr isochrones to estimate where the binary sequence would lie for different mass ratio ($q$) values.
    In these filters, the binaries with $q \gtrsim 0.5$ are largely distinct from the main sequence and easy to isolate.
    To measure the binary fraction in each $q$ range, we draw a box bounded by these isochrones and compare binary counts against main-sequence stars within the same color range.
		\label{F:binaries}}
\end{figure}

We make an observationally motivated estimate of the binary population in 47 Tuc.
On a color-magnitude diagram (CMD) the binary sequence lies above the main sequence (brighter and redder), and the extent of this offset depends on the mass ratio ($q$) of the binary pair.  A system with a mass ratio of $q=1$ has two identical stars and simply twice the flux in every filter, leading to a vertical rise of $\sim$0.75 mag and no shift in color.  
Systems with $q<1$ will be brightened by a lesser amount and reddened to a small degree depending on the specific $q$ value.
While the UV filters mentioned in Section~\ref{ss:PMs}
provide excellent positional information and allow for high-quality proper motion determination, they unfortunately are not ideal for identifying binary systems.  
In the UV CMD, the binary population is largely blended with the broad scatter of the main sequence, likely due to multiple populations with differing compositions \citep{Richer_etal_2013}.
Only those binaries with a mass ratio close to unity stand significantly apart from the main sequence, and even those have substantial contamination.  
Figure~\ref{F:binaries} shows a CMD of 47 Tuc from another data set using the redder filters F606W and F814W \citep[][GO-10775]{Sarajedini_etal_2007}.
These optical/NIR data cover a similar field of view (central $100''$) and provide a clearer distinction between the main sequence and the binary sequence, allowing us to distinguish systems down to $q\approx 0.5$.
The separation of the binaries from the main sequence occurs because the choice of filters can have a large impact on the morphology of the CMD. Both a small scatter and a shallower slope of the main sequence lead to a more distinct separation of binary systems from the main sequence.
For the low-$q$ binaries that remain hidden within the spread of the main sequence ($q \lesssim 0.5$), the large majority of them have total binary masses below $1 M_\odot$. 
Their masses will cause them to follow the distributions of similar-mass main-sequence stars and are thus left to be included as part of the general low-mass population.

For a more refined characterization of the binaries, we look at the binary fraction in three different $q$ ranges. 
We find that for the ranges $0.5<q<0.7$, $0.7<q<0.9$, and $q>0.9$, the binary fractions equal 0.0265, 0.0098, and 0.0039, respectively.
These are displayed in Figure~\ref{F:binaries}, and the specific values are used in computing the binary mass distribution in Section~\ref{sS:binary_params}. 

The isochrones used to calculate stellar masses were built from MESA stellar evolution models \citep[Modules for Experiments in Stellar Astrophysics;][]{PaxtonEtal2011,PaxtonEtal2013,PaxtonEtal2015} with a metallicity $z=0.003$ and age 11 Gyr.  The fluxes in each filter were determined using Phoenix atmospheres\footnote{\url{https://phoenix.ens-lyon.fr/Grids/BT-Settl/CIFIST2011\_2015/}}, by \citet{BaraffeEtal2015} and \citet{Allard2016}.


\subsubsection{Stellar Remnants}\label{sS:Remnants}

In addition to the concentrated binary systems, 47 Tuc harbors a collection of dark stellar remnants in the form of heavy WDs, NSs, and sBHs.  Most of these objects are virtually impossible to directly detect, but we can infer their parameters based on an initial mass function (IMF) for the cluster.

\begin{figure}[t]
	\centering
	\includegraphics[width=0.47\textwidth]{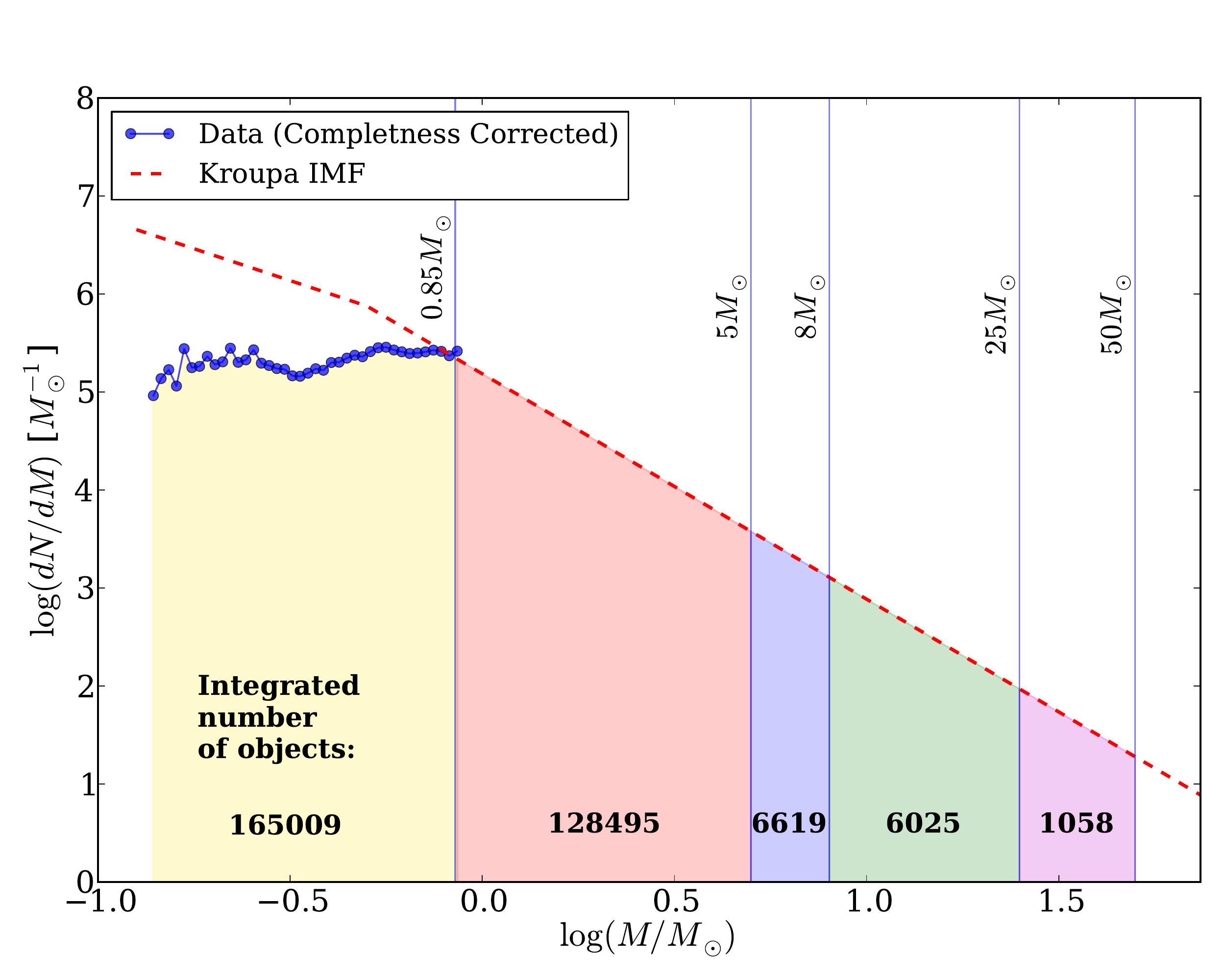}
	\caption{A Kroupa initial mass function is used to estimate the number of progenitor objects that form remnants of various masses.  
    The yellow region indicates the objects observable in the data.  They truncate at roughly $0.85M_\odot$, which is the turn-off mass of 47 Tuc.
    The red, blue, green, and purple regions, respectively, count progenitor objects that would create low-mass WDs ($<1M_\odot$), high-mass WDs ($>1M_\odot$), NSs, and sBHs.  For the species that end up above $1M_\odot$ (blue, green, purple), we use these counts to infer their population parameters and include them in the velocity dispersion model.
    The large number of predicted low-mass WD progenitors does agree with observed WD counts when one considers the WD birth rate, the time a WD takes to cool below detection thresholds, and the age of the cluster \citep{GoldsburyEtal2016}.
    Completeness corrections were done with artificial star tests \citep[also see][]{GoldsburyEtal2016}
		\label{F:mass_function}}
\end{figure}

In Figure~\ref{F:mass_function} we have plotted the observed stellar-mass function of the visual/NIR data as blue points.  Overlaid as a dashed red line is a \citet{Kroupa_2001} IMF normalized to the heaviest observed objects. 
The Kroupa IMF follows a broken power law, $N\propto M^{-\alpha}$, where $\alpha = 0.3, 1.3,$ and $2.3$ in the mass ranges $M<0.08$, $0.08<M<0.5$, and $M>0.5\ M_\odot$, respectively.
The deficit of low-mass stars between the data and IMF is presumably due to the preferential loss of low-mass stars from our field of view (segregation) and from the cluster as a whole (evaporation).
\citet{Baumgardt_Sollima_2017} verified this strong depletion of low-mass stars, and the present-day mass function agrees well with an evolved Kroupa IMF \citep{Baumgardt_Hilker_2018}.
We make the assumption that the highest mass stars still observed in the cluster ($\sim 0.85\ M_\odot$) remain at their natal abundance and use this IMF to estimate how many progenitor objects formed at different masses.
This is, in several ways, a conservative approach to estimating the number of stellar remnants.  The Kroupa IMF is among the steepest mass functions, and assuming some fraction of the $0.85\ M_\odot$ stars have been lost would only serve to raise the IMF and increase the remnant estimates.
By keeping our estimates conservative the final predicted IMBH mass is likely to be an overestimate, as there is potentially more remnant mass that we have not accounted for.

As mentioned previously, we are interested in tallying objects that have a mass above $1.0\ M_\odot$, and would therefore be distributed more centrally than main-sequence stars.  The red region of Figure~\ref{F:mass_function} marks progenitor stars that would form WDs of $<1.0\ M_\odot$. These WDs would relax into a distribution comparable to the main-sequence stars.  The blue region marks progenitors that produced WDs of mass $1.0\ M_\odot < M < 1.4\ M_\odot$ \citep{Cummings_etal_2016}.
The green region indicates the more massive progenitors that form NSs,
and the purple region marks the progenitors that produce sBHs of up to $10\ M_\odot$ \citep{Spera_Mapelli_2017}.
The number of objects in each colored mass range is determined by integrating the mass function.
We cut off the IMF beyond $M=50\ M_\odot$ because the BHs above $10\ M_\odot$ tend to be completely lost during cluster evolution \citep{Morscher_etal_2015}.

We have to consider that natal kicks imparted upon formation as well as gravitational encounters within the cluster will cause the loss of NSs and sBHs.  The retention fraction in globular clusters is not a well-constrained quantity.  Simulations tend to produce values anywhere in the range of about $5-50\%$ \citep{Mackey_etal_2008,Moody_Sigurdsson_2009,Morscher_etal_2015,Baumgardt_Sollima_2017}.
We adopt a value of 8.5\% retention for NSs and sBHs.  This choice will be explained in Section~\ref{sS:results_retention} where we also explore a range of retention values and their effect on the velocity dispersion profile.
Building the mass distributions of these heavy remnant objects is explained in the context of the velocity dispersion model in Section~\ref{sS:remnant_params}.

\section{An Analytic Model for Velocity Dispersion}\label{S:Model}

\subsection{Jeans Model}\label{sS:Jeans}

In order to determine whether or not our observations are consistent with a central IMBH, we create a theoretical velocity dispersion model that includes
the contributions of all the non-IMBH cluster objects as well as an IMBH mass parameter that affects the velocity dispersion signal in the core.
We use a 3D version of the King density template \citep{King_1962} as a starting point in determining the shape of the velocity dispersion profile.  
The density distribution,
\begin{equation}\label{E:rho_king}
	\rho(r) = K \left[  \frac{1}{(r^2 + a^2)^{3/2}} - \frac{1}{(r_\text{t}^2 + a^2)^{3/2}} \right], \ (r \leq r_\text{t}),
\end{equation}
describes how the density ($\rho$) falls off with 3D radius ($r$) as we move away from the core.  The density drops to zero at a prescribed tidal radius ($r_\text{t}$) which is set by the Galactic potential.  
We adopt a fixed tidal radius of $r_\text{t}=42'$ \citep[][2010 edition]{Harris_1996} for all populations discussed in this paper.
The profile scales with $K$, a constant that encodes the total mass contained in the distribution, and the parameter $a$ describes an effective core radius.  For radii interior to $a$ the distribution is largely flat.

The density profile and the associated enclosed mass as a function of radius, $M(r)$, can be used in the isotropic Jeans equation,
\begin{equation}\label{E:jeans}
	\frac{d}{dr}(\rho \sigma^2) = -\rho \frac{d\Phi}{dr} = -\rho \frac{G}{r^2} M(r),
\end{equation}
to isolate the square of the velocity dispersion $\sigma^2(r)$.  Here $\Phi$ and $G$ are the gravitational potential and constant, respectively. 
We use the non-rotating and isotropic version of the Jeans equation as the degree of rotation and anisotropy is very low in the core region of 47 Tuc where our proper motion data occur \citep{Heyl_etal_2017,Watkins_etal_2015,Bellini_etal_2017}.
The squared velocity dispersion then takes the form,
\begin{equation}\label{E:sig2r}
	\sigma^2(r) = -\frac{G}{\rho(r)}  \int_0^r   \frac{\rho(r)M(r)}{r^2} dr.
\end{equation}
This, of course, is the velocity dispersion profile for the 3D radius $r$.
Once $\sigma^2(r)$ is obtained, 
it is integrated against the density along the line 
of sight to produce the projected squared velocity dispersion,
\begin{equation}\label{E:sig2R}
\begin{split}
	\sigma_p^2(R) =& \frac{\int_R^\infty \sigma^2(r)\rho(r) r (r^2-R^2)^{-1/2} dr}    {\int_R^\infty \rho(r)  r (r^2-R^2)^{-1/2} dr} \\
    =& M_\text{T}F(R),
\end{split}
\end{equation}
where $R$ is the on-sky projected radius from the center.  
In the second line, we have simply parceled the equation into two parts: a function that holds the $R$ dependence and a scaling term out front. $F(R)$ will have units of
(velocity$^2$ mass$^{-1}$)
and describes how the velocity dispersion falls off with $R$.  The $M_\text{T}$ term is the total mass of the distribution in question and scales the magnitude of the velocity dispersion profile.  It arises from $M(r)$ in Equation~\ref{E:sig2r}, which involves an integral over the density profile and a coefficient that can be written in terms of $M_\text{T}$.  We adopt this format for ease in writing separate component contributions later on in Section~\ref{sS:remnant_params}.

The operations between 
Equations~\ref{E:jeans} and \ref{E:sig2R} are fully generic manipulations of the Jeans equation and make no specifications about the density profile.
In our full density distribution, we combine different King templates with individual total masses and core radii to capture the unique dynamical contributions of the various cluster populations.
For an analogous procedure using Gaussian distributions and with detailed formalism, see \citet{EmsellemEtal1994} and \citet{Capellari2008}.
Before we apply the Jeans equation directly to our cluster model, there are several nuances to consider which we describe in the following subsections.

\subsection{Treatment of Core Parameter $a$}\label{sS:aCl_aPM}

The parameter $a$ in Equation~\ref{E:rho_king} acts as a characteristic core radius.  However, when looking at different mass groups, there is not a single global value for $a$ because objects of different masses will be distributed with different core radii.  
It makes more sense to describe populations of different masses using individual $a$ values.
To correct for the inevitability that our tracer stars (those for which we have proper motions) are not perfectly representative of the true mass distribution, we make a distinction between two different core radius parameters,
$a_\text{Cl}$ and $a_\text{PM}$.  

In the right-hand side of Equation~\ref{E:jeans}, the factor $-d\Phi/dr = -GM(r)/r^2$ generates the force acting on each particle.  The $M(r)$ in this part of the equation must therefore reflect the true underlying mass distribution of the cluster, regardless of whether or not those objects are detected in the photometry.
Calculating this $M(r)$ term thus requires using a core radius parameter that encompasses all masses, even objects below our detection threshold. This is our $a_\text{Cl}$.  
It is an average core radius for all components of the cluster and is left as a free fitting parameter given its role of encompassing both detected and undetected mass.

Conversely, the $\rho$ terms of Equation~\ref{E:jeans} describe the distribution of tracer particles whose motions we are observing.  These tracers need not be the same objects that are generating the full potential.  We therefore calculate the remaining $\rho(r)$ terms 
(in Eq.~\ref{E:jeans}-\ref{E:sig2R})
using the measured core radius of the stars for which we have proper motions, $a_\text{PM}$.  
This $a_\text{PM}$ is a fixed quantity with a value measured by conducting a likelihood fit of the tracer particle positions to the King density template of Equation~\ref{E:jeans}.
In doing so we find $a_\text{PM}=36.0''$. 

Because our tracer stars (even with completeness corrections) are likely somewhat biased toward higher mass stars, we expect a difference between $a_\text{Cl}$ and $a_\text{PM}$.  This formulation allows $a_\text{Cl}$ and $a_\text{PM}$ the freedom to differ.

While $a_\text{PM}$ is calculated by fitting stellar positions, the free parameter $a_\text{Cl}$ is fit using the velocity information of the tracer stars.  This is described in more detail in later sections.

The binary and stellar remnant populations that we discuss in the next subsections will also have their own distribution parameters.
Due to the differing concentrations, the functional form of the velocity dispersion in Equation~\ref{E:sig2R} now becomes $F(R|a)$, where $a$ is specific to the mass distribution in question.

\subsection{Inclusion of Binary Populations}\label{sS:binary_params}

If there is no IMBH in the mass distribution, we expect the velocity dispersion to flatten off in the core as $M(r)\rightarrow 0$.  However, a central BH point mass causes $M(r)$ to remain positive for any $r>0$ and produces a rising slope into the cluster core.  A concentrated population of objects in the cluster can also cause the velocity dispersion profile to rise farther into the core before flattening out, mimicking the central effects of a BH.  When testing for an IMBH, it is therefore important to take careful consideration of populations more centrally concentrated than the typical main-sequence star, which may cause the velocity dispersion to rise deeper into the core.

For this reason, we model the binary population separately from the general lower mass cluster stars.  However, the binary systems are not a single homogeneous population in terms of their mass.
For a given primary mass (i.e. position along the main sequence) systems with high mass ratios are more massive than those with low $q$-values, and for a given mass ratio, binaries at the top of the main sequence are more massive than those below them.  
The higher mass systems will have correspondingly small distribution parameters.  We therefore break the binary contribution into components to account for binary subpopulations with different distributions.  
The expression for the squared velocity dispersion of the binary stars becomes
\begin{equation}\label{E:sig2R_expand}
\begin{split}
    \sigma_p^2(R) = \sum_{j,\alpha} M_\text{bin}^{j,\alpha} F(R|a_\text{bin}^{j,\alpha}).
\end{split}
\end{equation}
The index $j$ signifies different mass bins down the main sequence in the range of $0.85$--$0.55 M_\odot$.
The index $\alpha$ specifies one of the three mass ratio ranges seen in Figure~\ref{F:binaries}: lower ($0.5<q<0.7$), moderate ($0.7<q<0.9$), or high ($q>0.9$).  
The mass of binaries in each $j$-bin and $\alpha$-range ($M_\text{bin}^{j,\alpha}$) is calculated as
\begin{equation}\label{E:Mbin_ialpha}
\begin{split}
	M_\text{bin}^{j,\alpha} =&  f^{\alpha}  \left(\frac{A_\text{MS}}{A_\text{bin}}\right) \left(1 + \langle q \rangle ^\alpha \right)  \left(\frac{M_\text{MS}^j}{M_\text{obs}}\right) M_\text{Cl}.
 \end{split}
\end{equation}
The first item, $f^\alpha$, is the observed binary fraction for a given mass ratio range ($\alpha$).  These values are determined by counting stars in a section of the CMD as seen in Figure~\ref{F:binaries} and Table~\ref{T:Model_inputs}.  We assume that these binary fraction values hold true for the entire main sequence that is within our sensitivity limit. 
The $A$ factors are the proportion of the population that is visible within the $100''$ field of view according to the distribution parameter for that population.  
Therefore, the term $A_\text{MS}/A_\text{bin}$ modifies the observed binary fraction to describe the entire cluster, including objects beyond the image field.  
This is necessary because the binary fraction is expected drop away from the core since the binary systems are more concentrated than the main-sequence stars.  The factor $(1 + \langle q \rangle ^\alpha)$ accounts for the mass of the secondary star in the binary system where $\langle q\rangle^\alpha$ is the midpoint of the mass ratio range $\alpha$.  
At the end of Equation~\ref{E:Mbin_ialpha}, the quantity $M_\text{MS}^j$ is the observed mass in main sequence bin $j$, obtained by counting stars and consulting the isochrone.  It is divided by $M_\text{obs}$, the total mass of the observed main sequence (again by consulting the isochrone), to determine the mass proportion of the entire cluster our selection $j$ corresponds to.  Finally, multiplying by $M_\text{Cl}$ scales this proportion to the fitted dynamical cluster mass.

Because of the shape and the width of the main-sequence and giant branch in Figure~\ref{F:binaries}, the binary stars cannot be identified for all masses, and thus, their distributions cannot be measured directly.  We make use of the results of \citet{GoldsburyEtal2013} to determine the $a^{j,\alpha}_\text{bin}$ parameter of Equation~\ref{E:sig2R_expand}. In order to quantify the mass segregation in 54 globular clusters (including 47 Tuc), \citet{GoldsburyEtal2013} measured the distributions of various stellar groups, determining a power-law relation between the objects' mass and their distribution parameter.
They fit a mass segregation power law of $R_c = A(M/M_\odot)^B$, finding best-fit values of $A=24.4\pm0.6$ and $B=-0.95\pm0.05$ for 47 Tuc.
In their paper, $R_c$ works out to be equivalent to our $a$ parameter.  We use this power-law to determine  $a^{j,\alpha}_\text{bin}$ given that the mass of the binary system is the mass of the primary times the factor $(1 + \langle q \rangle ^\alpha)$.

\subsection{Inclusion of Dark Stellar Remnants}\label{sS:remnant_params}

The dark stellar remnants are included in a similar but simpler manner.  Each of the heavy WD, NS, and sBH populations are given a single distribution specified by their total mass and core radius parameter.  
Having no direct measurements of the dark remnant populations, we are required to make
some simplifying assumptions.
The heavy WDs are assumed to have a mean mass of $1.2\ M_\odot$, the midpoint of the considered mass range ($1\ M_\odot$ to the Chandrasekhar limit $1.4\ M_\odot$).  The NSs are all conservatively estimated to have masses of $1.4\ M_\odot$, and the sBHs are all assumed to have masses of $10\ M_\odot$
\citep{Morscher_etal_2015,Cummings_etal_2016,Spera_Mapelli_2017}.  
The populations' overall masses are simply given by the predicted number of objects multiplied by their mass.  
The number of objects in each of Figure~\ref{F:mass_function}'s colored mass ranges is determined by integrating the mass function.  This value is then corrected to include everything beyond the image field of view.  This correction process is analogous to the treatment used for binaries within Equation~\ref{E:Mbin_ialpha}.  Using estimates of the remnant masses,
we can infer their distributions and include them in our velocity dispersion model using the \citet{GoldsburyEtal2013} mass segregation power law as was done for the binary systems.
However, the sBHs and NSs will suffer additional losses, largely due to natal kicks from their progenitor supernovae.  We assume a moderate value of 8.5\% retention for NSs and sBHs.
The projected squared velocity dispersion contribution from each dark stellar remnant populations becomes
\begin{equation}
\begin{split}
	\sigma^2_{p,k}(R) =& M_{k}F(R|a_{k}),
\end{split}
\end{equation}
where $k$ is an index that refers to the dark stellar remnant in question (WD, NS, sBH).
The full and final expression for the squared velocity dispersion comes together as
\begin{equation}\label{E:sig2R_full}
\begin{split}
	\sigma_p^2(R)  &=  M_\text{IMBH} F(R) \\
    			 &+ M_\text{Cl} F(R|a_\text{Cl}) \\
    			 &+ \sum_{j,\alpha} M_\text{bin}^{j,\alpha} F(R|a_\text{bin}^{j,\alpha}) \\
                 &+ \sum_{k} M_{k}F(R|a_{k}).\\
\end{split}
\end{equation}
In each case, the $F$ function encapsulates the radial dependence of the velocity dispersion for the component in question and is scaled by the total mass $M$ term of that component.
The IMBH term is slightly different in that the enclosed mass function ($M(r)$ of Equations~\ref{E:jeans} and \ref{E:sig2r}) is simply the constant $M_\text{IMBH}$ because we have considered it a point mass centered on $r=0$.  It therefore is influenced only by the $a_\text{PM}$ value within $\rho(r)$. It has no distribution parameter equivalent to the $a_\text{Cl},a_\text{bin}$, or $a_{k}$ found in the other terms.

We have chosen to split the heavy objects by their type rather than their mass group simply for ease of estimation.  One could also imagine lumping all objects of similar mass regardless of whether they are binary, WD, NS, or sBH.  We found it simpler to estimate object counts by considering direct observational evidence of the binary sequence and estimating progenitor counts for the remnants.

We have now built a $\sigma_p^2(R)$ distribution up from 47 Tuc's different mass components.  Using measured binary fractions and inferred distributions, we account for the individual contributions of binary subpopulations that have differing distributions and total masses.  An IMF normalized to observed cluster stars allows us to estimate how many dark stellar remnants remain in the cluster.  The fit parameters $M_\text{Cl}$ and $a_\text{Cl}$ describe the gross shape of the velocity dispersion curve (its vertical scale and the radius at which it falls off), while the concentrated binaries and remnants bring up the velocity dispersion value toward the core.  Any central rise observed in the data that is not due to the concentrated populations is taken up by the IMBH fit parameter $M_\text{IMBH}$.

Putting all of this together produces a radial profile for the projected velocity dispersion
that depends on three parameters: $M_\text{IMBH}$, $M_\text{Cl}$, and $a_\text{Cl}$.  The rest of the inputs are measured or inferred quantities.
The projected velocity dispersion profile, $\sigma_p(R | M_\text{IMBH}, M_\text{Cl}, a_\text{Cl})$, is now in a form that can be fit to the data and the optimal values for the three fitting parameters found.

\subsection{Binned vs. Unbinned Analysis}\label{sS:unbin}

With a theoretical velocity dispersion model now built, we are in a position to find the set of the three parameters ($M_\text{IMBH}$, $M_\text{Cl}$, and $a_\text{Cl}$) that best 
fit the observations. This is most simply and commonly done by radially binning the data, computing the velocity dispersion in each bin,
then finding a least-squares fit of the model to the data.

We must, however, be cautious of trusting the results of a binned data set.  
Any binning scheme has some degree of arbitrariness 
that may influence the result in unpredictable
ways.  At the very least, introducing bins will smear out the radius-dependent signal 
over a larger range of radii.  At worst, the choice of binning system may
unintentionally amplify or suppress a particular result.

\begin{figure}[t]
	\centering
	\includegraphics[width=0.48\textwidth]
    {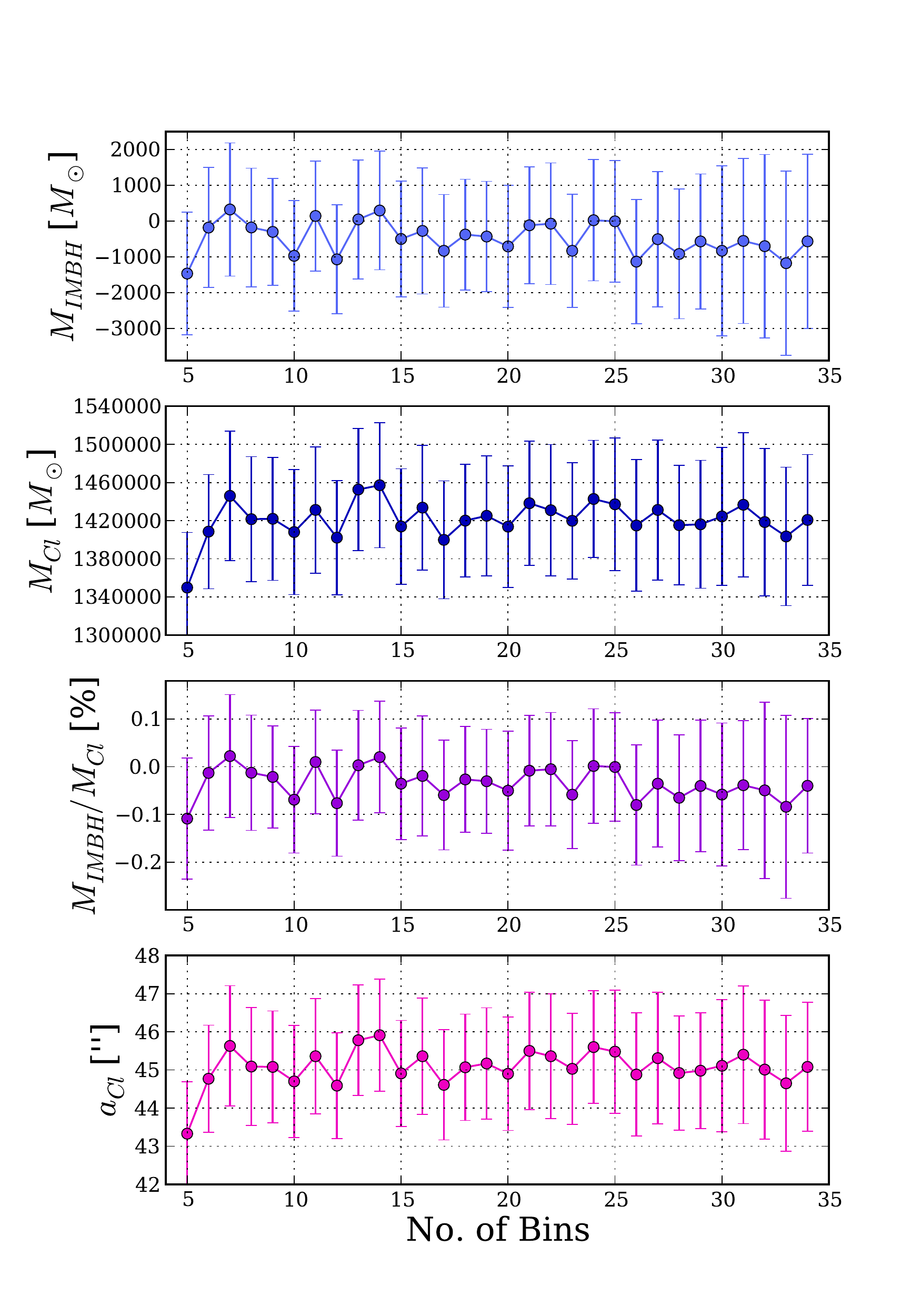}
    \caption{
     To test the influence of binning scheme on the fit results a radial binning scheme is chosen that follows a logarithmic spacing out to the limit of the UV data ($120''$).  When the number of bins (and thus, the size of the bins) is changed, the binned-fit parameters fluctuate within their uncertainty levels.
    Note: A negative IMBH mass simply means the non-IMBH components of the model produce a higher velocity dispersion in the core than the observations indicate.
 		\label{F:bin_effects}}
\end{figure}

As a test of the influence of bin choice, we set up a logarithmically spaced binning scheme then steadily change the number (and thus the size) of the bins.
Figure~\ref{F:bin_effects} shows how the binned-fit parameters fluctuate as we slowly scale the bin sizes.  
We find that the jitter the parameters experience is generally within the typical uncertainty of each parameter.
Despite the agreement between binning schemes, we still decide to use an unbinned fitting method for two reasons: an unbinned system is more inclusive of every piece of information available to us, and it also removes any human element of bin choice.
The mean values of each of the Figure~\ref{F:binaries} parameters are slightly offset from the unbinned results reported later in Section~\ref{S:Results}.  We suspect this is due to the smearing-out of the data when employing bins, though the different methods do agree within their respective uncertainties.
We use the following unbinned likelihood maximization technique to find the best-fit velocity dispersion model.

Under the assumption of isotropy in the core for 47 Tuc,
we assume the observed velocity components ($v_x$, $v_y$) of each star are samples from Gaussian-distributed populations centered on zero.  The widths of these Gaussian populations are estimated by the model velocity dispersion value at each star's projected distance from the core, broadened by the uncertainty in the velocity measurements. We find the set of 
$M_\text{IMBH}$, $M_\text{Cl}$, and $a_\text{Cl}$ parameters for $\sigma_p(R)$ that maximize the global log-likelihood function,
\begin{equation}\label{E:lnL}
\begin{split}
	\ln L =& \sum_i  - \frac{v_{x,i}^2}{2(\sigma^2_{p,i} + \delta^2_{x,i})}  -  \frac{v_{y,i}^2}{2(\sigma^2_{p,i} + \delta^2_{y,i})}  \\
	 &- \frac{1}{2}\ln(\sigma^2_{p,i} + \delta^2_{x,i}) - \frac{1}{2}\ln(\sigma^2_{p,i} + \delta^2_{y,i}).
\end{split}
\end{equation}
Here, $\sigma_{p,i}=\sigma_p(R_i|M_\text{IMBH},M_\text{Cl},a_\text{Cl})$ is the model velocity dispersion value at the projected radius of star $i$, and $\delta_{x,i}$ and $\delta_{y,i}$ are the measurement uncertainties in 
the velocity components of star $i$ that arise from our photometric data.  
This likelihood estimator is equivalent to Equation (8) of \citet{Walker_etal_2006}, but has been adapted for 2D proper motions and a continuous $\sigma_p^2(R)$ rather than 1D radial velocities and binned velocity dispersion values.  Furthermore, we ignore the constant offset introduced to the $\ln L$ value by Gaussian normalization.
For a full derivation and explanation of the entire unbinned likelihood  technique, see \citet{GoldsburyEtal2016}.
The specific parameters that went into our model are shown in Table~\ref{T:Model_inputs} and the resulting velocity dispersion profiles are plotted in Figure~\ref{F:VD}.

We should note that this model is not restricted from fitting a negative $M_\text{IMBH}$ parameter and clarify what a negative value indicates.
The model cannot measure mass directly, but only the amount of dispersion
present in the stellar velocities.  A negative IMBH mass, while clearly an unphysical
phenomenon if taken at face value, simply indicates that the chosen components of the model
(e.g. Table~\ref{T:Model_inputs}) produce a higher velocity dispersion in the core than
is seen in the data.  The fitting process adopts a negative IMBH mass parameter in an attempt
to bring down the velocity dispersion in the core to improve the overall fit.  
This indicates that some imperfect assumptions and estimations were made in building the model and the negative IMBH mass is a sign that the model places more mass in the core than can be explained by the observed velocities.

\begin{table*}[t]  
\begin{centering}
\caption{Model inputs for binaries and stellar remnants } \label{T:Model_inputs}
\begin{tabular}{ p{1.7cm}p{2.3cm}p{2.0cm}p{1.0cm}p{1.2cm}p{1.2cm}p{1.5cm}p{1.0cm} }
\hline \hline
Object Type	& Subgroup Parameters/Type& Primary Mass ($M_p$) Range	& $\langle M_p\rangle$ & Object Mass& Retention Fraction& Population Mass & \ $a$ \\
		& 		 &   ($M_\odot$)  & ($M_\odot$)  &	($M_\odot$)   &	   &($M_\odot$)	  &	( $''$ )   \\
\hline \hline
Binaries&$q=[0.5,0.7]$              &0.80--0.86     &	0.83 &	1.33 &	1   &	7102 &	18.61 \\
        &($\langle q\rangle=0.60$)  &0.74--0.80     &	0.77 &	1.23 &	1   &	6537 &	20.07 \\
		&($f=0.00265)$		        &0.67--0.74     &	0.71 &	1.13 &	1   &	5979 &	21.72 \\
		&		                    &0.61--0.67     &  	0.64 &	1.03 &	1   &	5110 &	23.71 \\
		&		                    &0.55--0.61     &	0.58 &	0.93 &	1   &	5011 &	26.12 \\

\\								
		&$q=[0.7,0.9]$              &0.80--0.86     &	0.83 &	1.50 &	1   &	2824 &	16.64 \\
		&($\langle q\rangle=0.80$)  &0.74--0.80     &	0.77 &	1.38 &	1   &	2592 &	17.95 \\
		&($f=0.00098)$ 		        &0.67--0.74     &	0.71 &	1.27 &	1   &	2363 &	19.42 \\
		& 		                    &0.61--0.67     &	0.64 &	1.16 &	1   &   2013 &	21.20 \\
		& 		                    &0.55--0.61     &	0.58 &	1.05 &	1   &	1965 &	23.36 \\

\\									
		&$q=[0.9,1.0]$              &0.80--0.86     &	0.83 &	1.62 &	1   &	1193 &	15.42 \\
		&($\langle q\rangle=0.95$)  &0.74--0.80     &	0.77 &	1.50 &	1   &	1094 &	16.63 \\
		&($f=0.00039)$		        &0.67--0.74     &	0.71 &	1.38 &	1   &	995 &	18.00 \\
		&		                    &0.61--0.67     &	0.64 &	1.26 &	1   &	845  &	19.65 \\
		&		                    &0.55--0.61     &	0.58 &	1.13 &	1   &	822  &	21.65 \\

\hline
Remnants& WD	                    & -- --         & -- --  & 1.2   &	1	    & 24000	 & 20.52 \\
		& NS	                    & -- --         & -- --  & 1.4   &	0.085	& 23800	 & 17.72 \\
		& sBH                       & -- --         & -- --  & 10    &	0.085	& 19000	 & 2.74 \\
\hline
\end{tabular}
\tablecomments{}{
	The object mass within a binary subgroup is calculated as $\langle M\rangle (1-\langle q\rangle)$.
    There is a total mass of $\sim 46,500 M_\odot$ in binaries and $\sim 27,600 M_\odot$ in stellar remnants (taking into account the 8.5\% retention of NSs and sBHs).
    Population masses for binaries are calculated as described in Equation~\ref{E:Mbin_ialpha}. Population masses for stellar remnants are calculated by assuming an object mass, determining the distribution parameter ($a$) from that mass, and then consulting the predicted progenitor counts of Figure~\ref{F:mass_function} that have been corrected to include the entire cluster instead of the limited field of view. Population mass values presented are before retention is considered.
    }
\end{centering}
\end{table*}



\section{Results}\label{S:Results}

\subsection{Model-Fitting Results}\label{Ss:fitting_results}

\begin{figure}[t]
	\centering
	\includegraphics[width=0.48\textwidth]{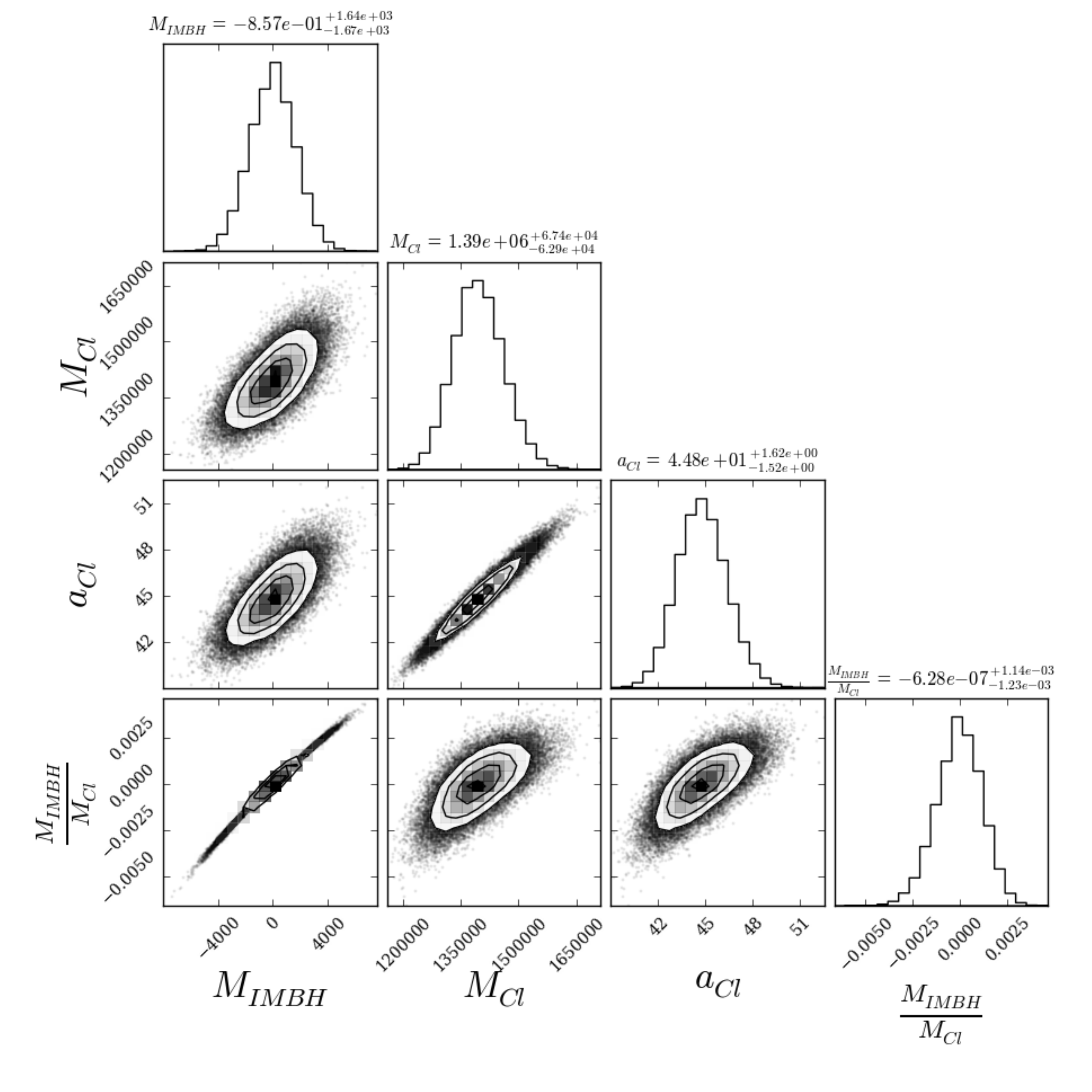}
	\caption{ 
       Results from bootstrapping the proper motion data and fitting
       the velocity dispersion model.  The bootstrap method 
       (randomly drawing stellar data points with replacement for each iteration) provides a spread of fit parameters
       around the best-fitting values that give a measure of the uncertainty 
       in each parameter.
		\label{F:param_corner}}
\end{figure}

Having built a velocity dispersion model that includes measured binary fractions, inferred dark stellar remnants, and an NS and sBH retention fraction of 8.5\%, we obtain the following best-fit parameter values: 
%
$M_\text{IMBH} =  40 \pm 1650 \ M_\odot$, 
$M_\text{Cl} = (1.39 \pm 0.07) \times 10^6\ M_\odot$,
and
$a_\text{Cl} = 44.8 \pm 1.6 ''$.
This provides us with an
$M_\text{IMBH}/M_\text{Cl}$ ratio of $0.00\% \pm 0.12 \%$.
Note that the binaries and remnants themselves sum to $\sim 5.3\%$ of $M_\text{Cl}$ (see Table~\ref{T:Model_inputs}) and are not included in the $M_\text{Cl}$ parameter.  Therefore, the entire mass of 47 Tuc comes to $1.47\times 10^6 M_\odot$.
%
%
%
%
%
The uncertainties in the fit parameters were determined through a bootstrapping method in which we
randomly sampled the stars (with replacement) to obtain a new data set to which we fit the
velocity dispersion model.  The spreads in the distributions of the resulting fit parameters provide
our estimates of uncertainty (shown in Figure~\ref{F:param_corner}).

\begin{figure}[t]
	\centering
	\includegraphics[width=0.48\textwidth]{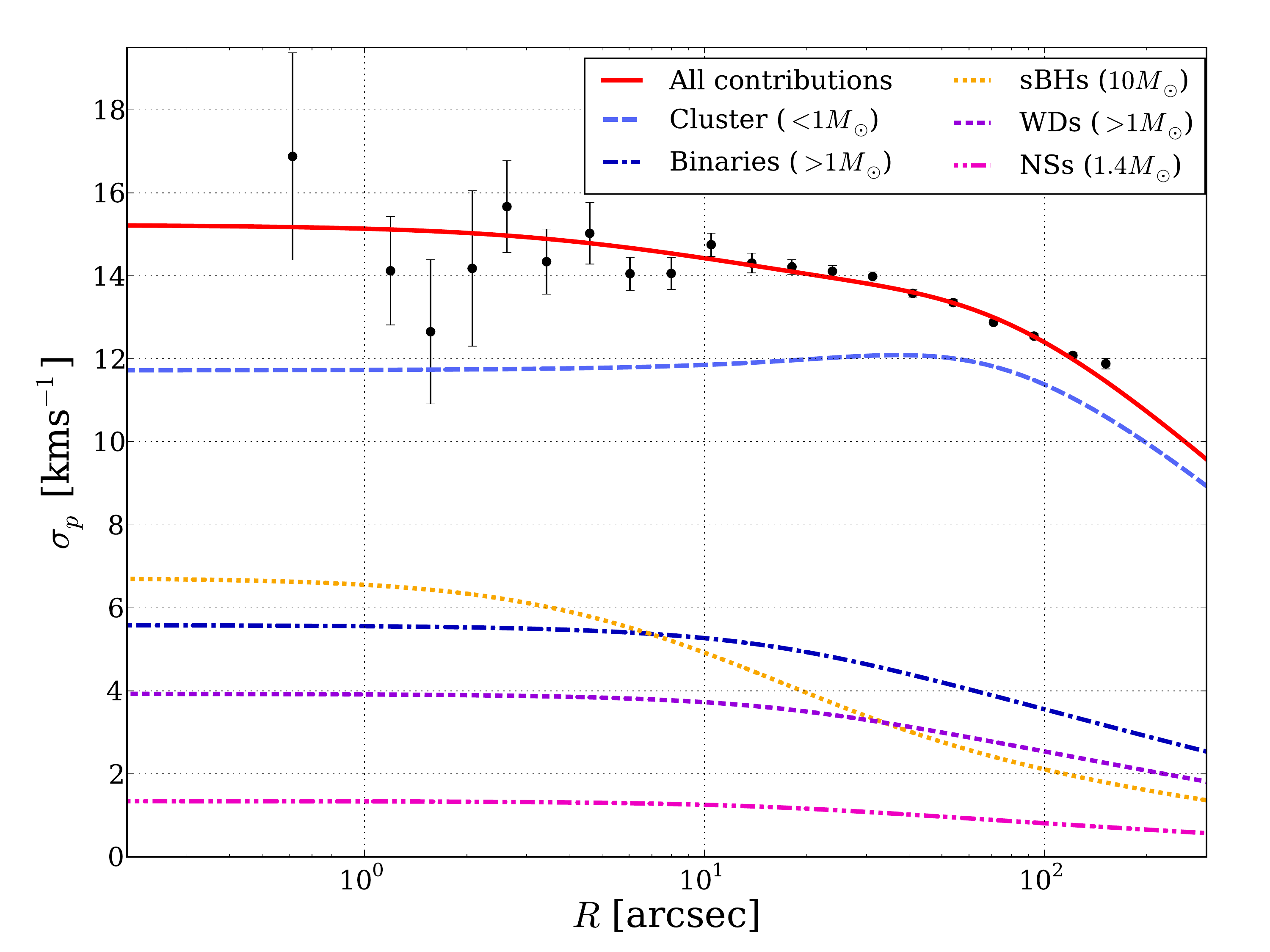}
	\caption{
    The solid red line shows the best-fitting (unbinned) velocity dispersion model.  
    It includes the IMBH, binary, remnant, and low-mass cluster object components.
    Model parameters are given in Table~\ref{T:Model_inputs}.
    Note that the black points display binned data and are included for visualization purposes only.  The model was fit using an unbinned likelihood maximization.  Dashed lines show how the different components
    (cluster stars, binaries, sBHs, WDs, and NSs) contribute to the overall velocity dispersion profile with the individual components being added in quadrature.  
    We see here that the sBHs and binaries 
    are the dominant contributors to a rise in the core.
		\label{F:VD}}
\end{figure}

The velocity dispersion model described by these best-fit parameters is shown in Figure~\ref{F:VD}, plotted over binned velocity dispersion data.  
The binned data are included for visualization purposes only as the fit was made using the unbinned likelihood maximization described in the previous section.
For concentrated heavy populations, the binaries and sBHs play dominant roles in the central velocity dispersion rise.  The binary population contains a large overall mass and their distribution causes the velocity dispersion to continue rising into $\sim 10''$.  The sBHs, while comprising a lesser total mass, are much more concentrated and continue this velocity dispersion rise into $\sim 2''$.  
The combined effect leaves no room for any IMBH that would drive the velocity dispersion curve even higher.

The results reported here are somewhat influenced by our choice of cluster distance.  
We have adopted a distance of $d=4.69$ kpc from \citet{Woodley_etal_2012}, who fit spectral energy distributions of 47 Tuc's WDs.
The choice of distance scales the observed velocities, raising or lowering the velocity dispersion profile.  This has a direct effect on the fitted cluster mass (set by the height of the profile) but has a less straightforward influence on the IMBH mass parameter, due to our inclusion of dark stellar remnants in the model.  The remnants were calculated independent of any velocity measurement, so they do not scale with distance.  Therefore, if we were to adopt a lower cluster distance (producing lower stellar velocities), the cluster mass would scale downward while the remnant population remains fixed.  This would make the contribution of the remnants proportionally more important in the core and further reduce the need for an IMBH influence.  

We verified this effect by considering the recent parallax distance of 4.45 kpc for 47 Tuc from \citet{Chen_etal_2018}.  Fitting the model with this lower distance dropped the cluster mass down to $M_\text{Cl}=1.20\times10^6\ M_\odot$ and the fitted IMBH mass to a negative value,  $M_\text{IMBH}=-560\ M_\odot$ (at 8.5\% sBH and NS retention).  
Our calculated stellar remnants already accounted for enough concentrated mass in the core.  A smaller cluster distance only serves to push the fitted IMBH mass lower.

Our fitted cluster mass is relatively large.  \citet{Giersz_Heggie_2011} and \citet{Lane_etal_2010} found 0.9 and 1.1$\times10^6\ M_\odot$, respectively.  The discrepancy could be influenced by the choice of distance as discussed above or potentially be due to our data limitations.  Our proper motion data only extend to  $120''$, a small fraction of the tidal radius,  $r_\text{t} \approx 42'$ \citep[][2010 edition]{Harris_1996}.  This means we are really only fitting the central region of the cluster.  Our cluster mass estimate involves integrating the entire density distribution and therefore assumes that the density beyond $120''$ follows a perfect King template out to the tidal radius. Deviation in the tails may contribute to errors in the cluster mass.

\subsection{Verifying model efficacy}

We would like to test our velocity dispersion model to ensure it is accurately representing
the data it is being fit to.  Being able to find a set of best-fit model parameters does not necessarily mean one has modeled the data well.
For example, any set of curved data will have a straight line that fits it best, but that does not mean a straight line is a particularly good description of those data.  
To determine the quality of our model, we employ a system of statistical resampling in order to assess the efficacy of our modeling process.  

\begin{figure}[t]
	\centering
	\includegraphics[width=0.48\textwidth]{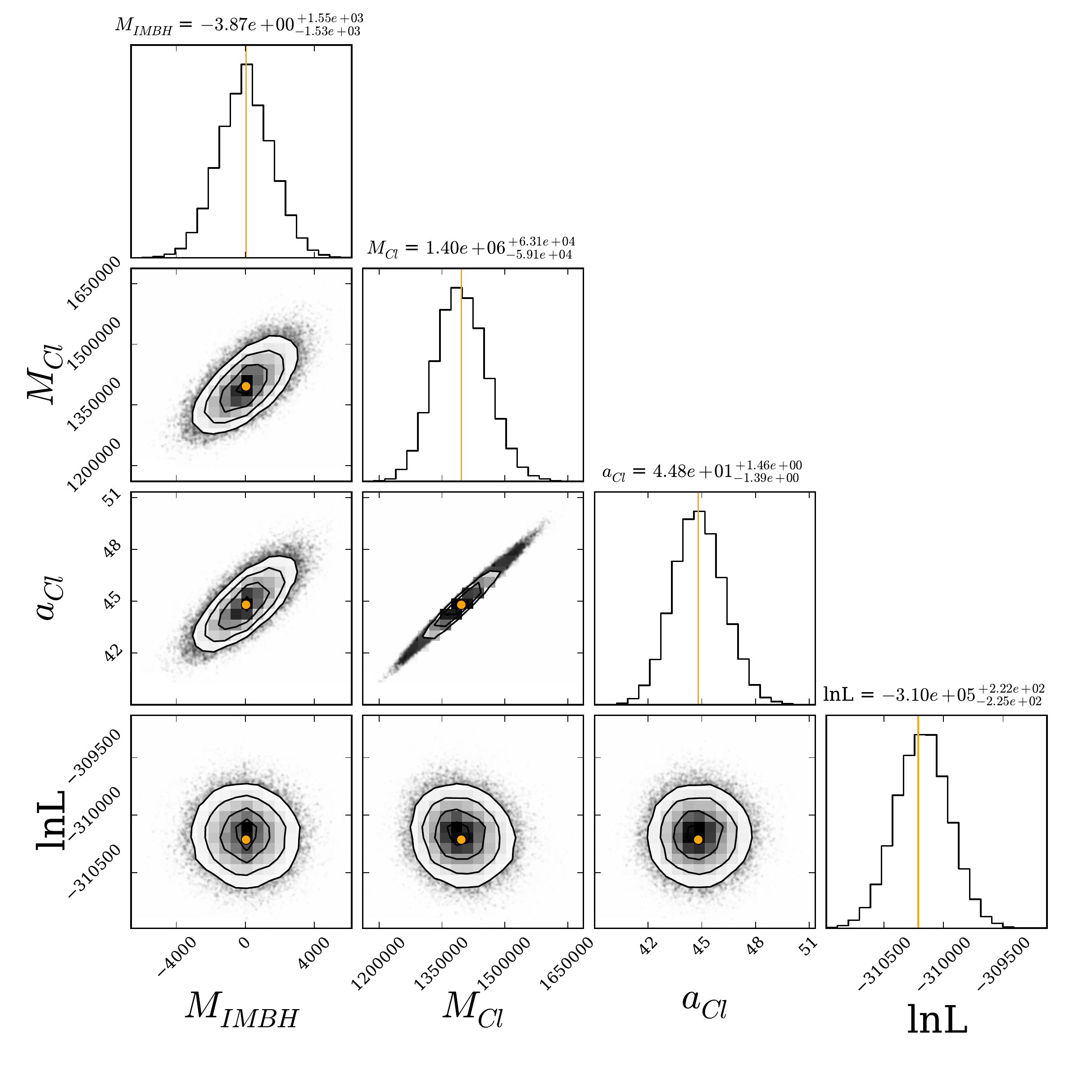}
	\caption{ 
    Presented here are the fit results of the artificially sampled data sets.
    The artificial velocities are drawn from the original data's best-fit velocity 
    dispersion model.  All fits used unbinned likelihood maximization.
   	The orange dots and vertical lines indicate the fit values of the original data.
    There are two major conclusions to be drawn from this exercise: (1) The proximity of
    each orange line to the mean of the $M_\text{IMBH},M_\text{Cl}$, and $a_\text{Cl}$ distributions
    indicates no systematic bias in our model's ability to recover input parameters, and
    (2) the proximity of the original data's $\ln L$ value to the mean of the artificial
    distribution shows our model is indeed a good fit to the shape of the velocity data.
		\label{F:BHlnLnewlnL}}
\end{figure}

We generate new $v_x$ and $v_y$ velocity components for each star.  These velocities are drawn randomly from Gaussian distributions whose widths are determined by the best-fit model velocity dispersion value at the radius of each data point and broadened by the uncertainty of the velocity measurement.  
We keep the original positions for each star to ensure the radial distribution remains unchanged, maintaining a consistent radial sampling in the core where there are few stars.
The velocity dispersion model is fit to this new set of artificial data, and the fit parameters as well as a goodness-of-fit measure, the log-likelihood ($\ln L$), are recorded. 
Repeating this process many times builds distributions of the fit parameters and $\ln L$ values.
The artificial data sets will clearly produce a Gaussian spread around the input model, but the pertinent question is whether or not the true data produce a similar fit.
Figure~\ref{F:BHlnLnewlnL} shows the distributions of the generated fit parameters and $\ln L$ values.  As can be seen, the orange lines/dots indicating the true values of the data fall very near the middle of each distribution.  
Such close agreement of the $M_\text{IMBH}, M_\text{Cl}$, and $a_\text{Cl}$ parameters tell us our model suffers no systematic bias in recovering input parameters.
The close agreement of the $\ln L$ value
indicates our model is a good morphological description of the data.
We can conclude that artificial velocity data generated from our model closely resemble the cluster's velocity data.

\subsection{Retention fraction of sBHs and NSs}\label{sS:results_retention}

\begin{figure}[t]
	\centering
	\includegraphics[width=0.45\textwidth]{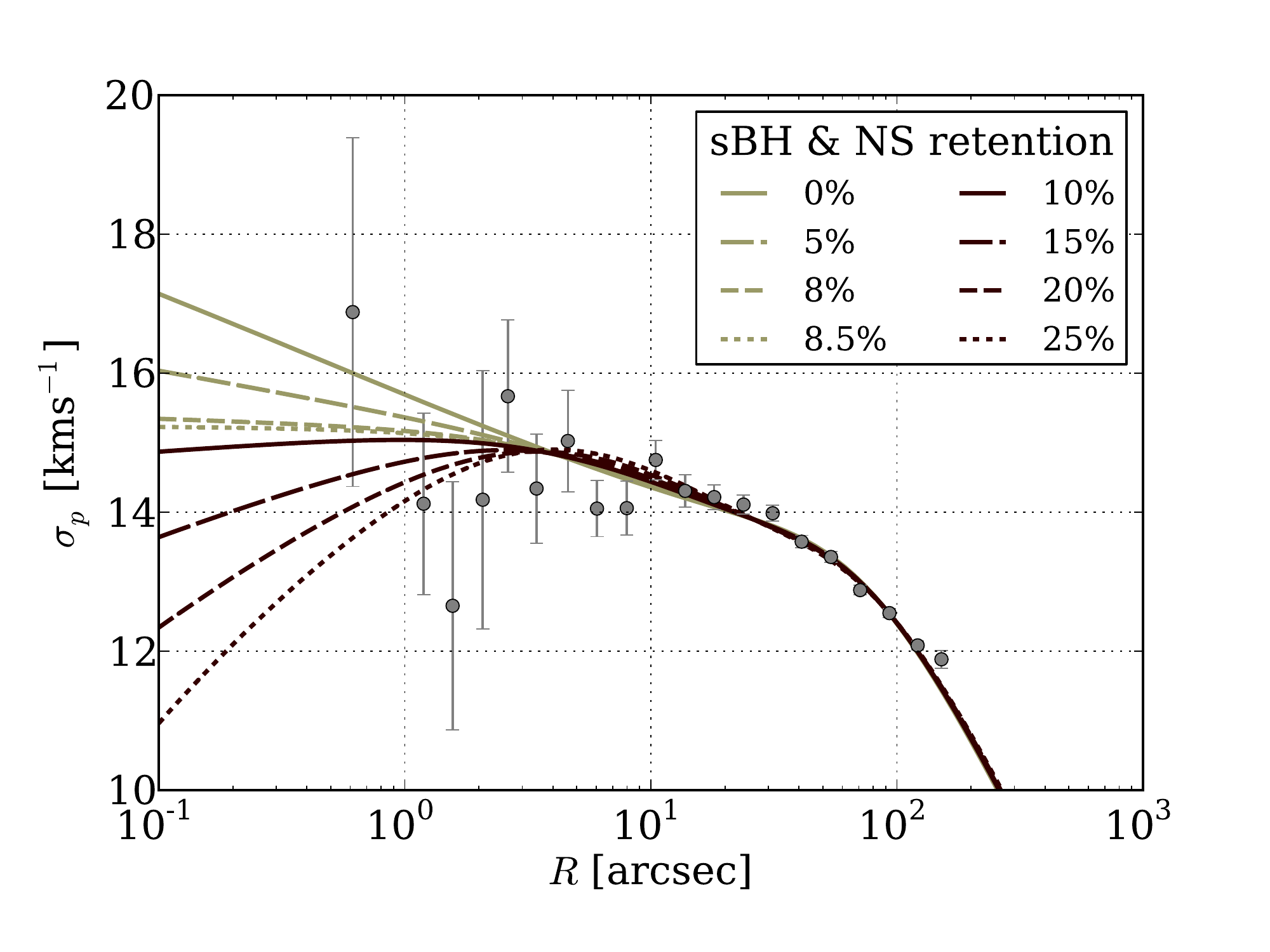}
	\caption{
    	Velocity dispersion profiles corresponding to the different NS and sBH
          retention fractions explored in Table~\ref{T:retention}. Retention values
          of $<8.5\%$ (light curves) show models that predict a positive IMBH mass.
          Retention values of $>8.5\%$ (dark curves) show models that predict a 
          negative IMBH mass.
          The overall log-likelihood difference between these models is small due to
          the relatively few data points at low radius where the models differ most
          substantially (see Table~\ref{T:retention}).
        \label{F:retention_profiles}}
\end{figure}

We explored a range of sBH retention rates to determine the effect it can have on the final fitted IMBH mass {(see Table~\ref{T:retention} and Figure~\ref{F:retention_profiles})}.
The negative IMBH models (dark curves of Figure~\ref{F:retention_profiles}) can be instructive in signifying inappropriate model assumptions, but they are ultimately unphysical systems. 
Our adoption of an 8.5\% retention fraction arises from choosing the highest likelihood model that produces nonnegative IMBH mass parameter (Table~\ref{T:retention}). 
There is a somewhat degenerate trade-off between the influence of tightly distributed massive objects versus a central IMBH. This trade-off happens in the $2''-10''$ region where the sBH distribution can produce a slope in the velocity dispersion.  Interior to $\sim2''$ nothing can produce a slope except for a point mass, so the IMBH term always dominates in this region.

We observe that in order to find the $M_\text{IMBH}/M_\text{Cl}$ fraction concluded by \citet{KiziltanEtal2017}, we would need to assume that essentially all $10\ M_\odot$ sBHs are ejected from the cluster. 
The only way we can see evidence of an IMBH at all is if we assume an sBH and NS retention fraction very close to zero.
Pushing the retention any higher than $\sim 8.5\%$ creates a larger velocity dispersion in the core than is observed, requiring a negative IMBH mass fit.
Interestingly, from the trials shown in Table~\ref{T:retention} and Figure~\ref{F:retention_profiles}, the $\ln L$ values' quadratic shape peaks near $18.4\%$ retention.  
This may indicate a model that overestimates the core contribution from remnant/binary populations (requiring negative $M_\text{IMBH}$) but fits the numerous data points at intermediate radii very well. 
We clearly see the trade-off between sBH populations and an IMBH that can produce similar dynamical effects.

\section{Conclusion}\label{S:Summary_Discussion}
\subsection{Summary}

Our use of UV imaging data has allowed us to glean velocity information from stars right into the very center of the globular cluster 47 Tucanae.
Historically, crowding has been a limiting factor in IMBH velocity dispersion searches as it severely reduces the numbers of stars measurable at very small projected radius.  Resolution and accuracy in this region are of key importance in trying to estimate the mass of a central IMBH through this method.
With visibility right in to the cluster core, we are able to probe the region of the cluster where an IMBH's influence would be most pronounced.

Beginning with a 3D King density template, we used the isotropic Jeans equation to build a velocity dispersion profile with the added influence of a central IMBH.  
The velocity dispersion profile was built with contributions from several subcomponents:
a central IMBH point mass,
a collective group of low-mass cluster objects ($\lesssim 1.0\ M_\odot$),
and individual contributions from the concentrated populations of the binary systems, heavy WDs, NSs, and sBHs.

We employed an unbinned likelihood analysis when fitting our model.  This technique
weighs the combined likelihoods of each individual star's velocities being drawn from a given velocity dispersion model.  It prevents the radial smearing-out that
occurs when data are binned together and removes any arbitrary human factor in choosing a binning scheme.

To determine how good a fit our velocity dispersion model is to the data, we randomly resampled the velocity components of our stars based on the best-fit velocity dispersion curve.  The log-likelihood of the real data's  best-fit model is in very good agreement
with the distribution of resampled log-likelihood values, indicating our model does a good job of describing the data.




Our analysis produces a best-fit value of $M_\text{IMBH} =40 \pm 1650 \ M_\odot$ 
for a central IMBH in the core of 47 Tucanae, giving an $M_\text{IMBH}/M_\text{Cl}$ ratio of $0.00\% \pm 0.12 \%$.
%
We find no evidence to suspect an IMBH is present in the core of 47 Tuc unless one also accepts a very low sBH and NS retention fraction.
Our result is consistent with zero and in contention with the results of \cite{KiziltanEtal2017}, who found an $M_\text{IMBH}/M_\text{Cl}$ ratio of $0.30\% ^{+0.20\%}_{-0.12\%}$.
We emphasize that the proper characterization of the binary population and sBHs can have a marked effect on the IMBH estimate and their omission will likely bias that estimate high.

\begin{table}[t]  
\begin{centering}
\caption{Exploring Retention Fraction}
\begin{tabular}{c c c c c c}
Retention   &  $M_\text{IMBH}$  &  $M_\text{Cl}$ &  $a_\text{Cl}$ & $\frac{M_\text{IMBH}}{M_\text{Cl}}$ & $\ln L$\\
\ (\%)	&	($M_\odot$) &	($M_\odot$)  & ( $''$ )	&  & \\
\hline \hline
	 0		&	4125 	&	1346143  &42.8	&	0.0031  & -310214.24\\
     5		&	1713 	&	1374272  &43.9	&	0.0012  & -310213.41\\
 	 8		&	 278 	&	1392757  &44.7	&	0.0002  & -310213.01\\
	8.5     &     40    &   1395974  &44.8  &   0.0000  & -310212.95\\
	10		&	-672 	&	1405850  &45.2	&  -0.0005  & -310212.78\\
	15		&  -3020 	&	1441640	 &46.6	&  -0.0021  & -310212.40\\
	20		&  -5317 	&	1482612  &48.1	&  -0.0036  & -310212.32\\
    25		&  -7545	&	1530150	 &49.8  &  -0.0049  & -310212.63\\
\hline
\end{tabular}
\tablecomments{}{
Enforcing various retention fractions for NSs and sBHs provides different
fit results.  Typical uncertainties for these parameters
are very similar to those reported in Section~\ref{Ss:fitting_results} (those results are the 8.5\% retention case shown here).
Retention below $\sim 8.5\%$ is needed to find a positive fit mass for an IMBH.  A retention
of essentially 0\% is required to obtain the \citet{KiziltanEtal2017} mass fraction of $0.30\%$.
It appears that retention
above $\sim 8.5\%$ starts to become incompatible with the core velocity dispersion seen in this 47 Tuc data. 
Corresponding velocity dispersion profiles are shown in Figure~\ref{F:retention_profiles}. }
\label{T:retention}
\end{centering}
\end{table}

\begin{figure}[t]
	\centering
	\includegraphics[width=0.45\textwidth]{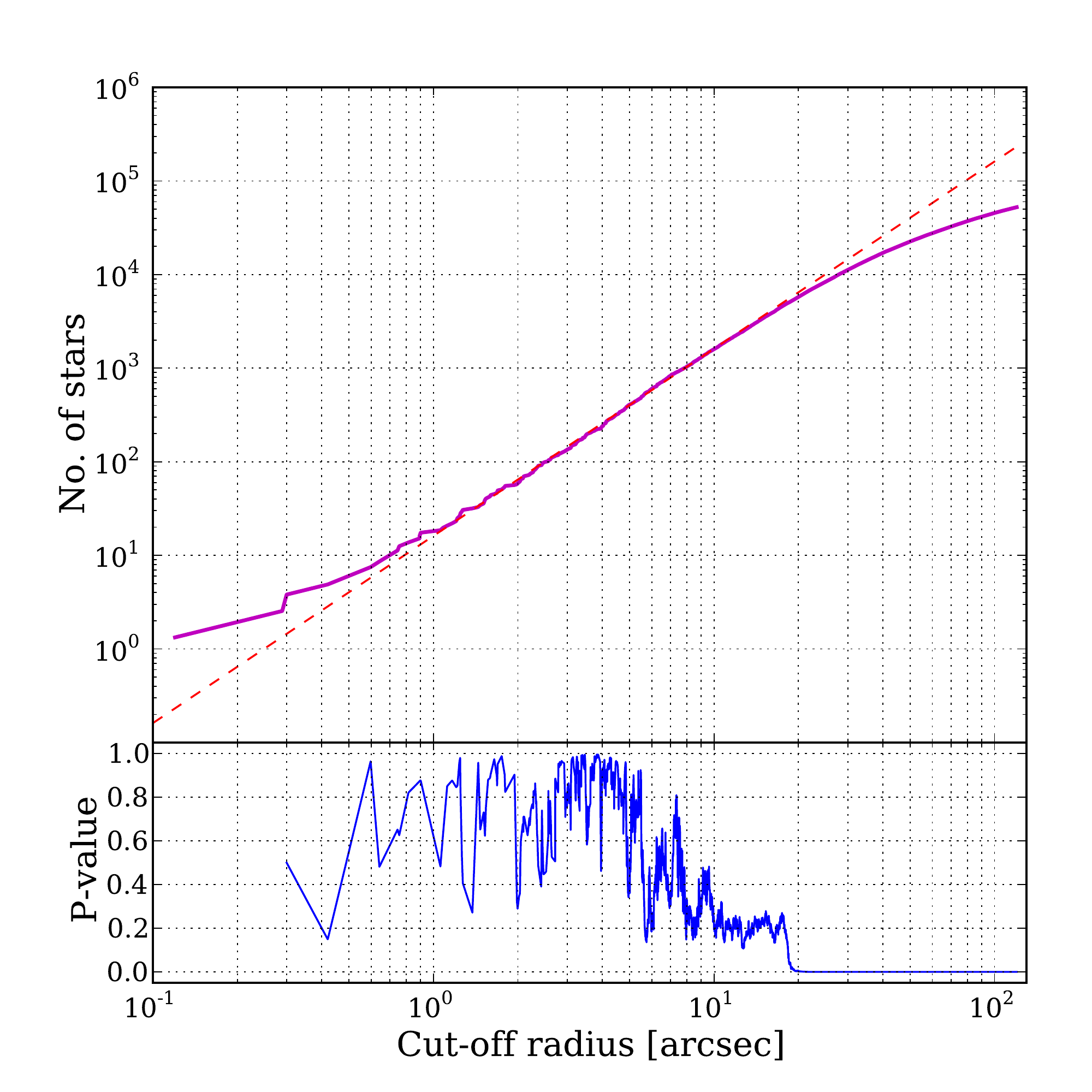}
	\caption{
    	The top panel shows the cumulative distribution of stars (incompleteness corrected) as a function of distance from the core.
    	At each point on the horizontal axis, we are considering all stars interior to that cutoff radius.
    	The dashed red line is an $R^2$ relation, indicative of a flat density distribution.  The apparent deviation in the core from the $R^2$ line is not statistically significant 
    	{as is evident in the bottom panel, which shows the $P$-value result of a Kolmogorov-Smirnov test.  It shows that we cannot reject the null hypothesis (i.e. that stars are drawn from an $R^2$ relation) to any high significance until $\sim 20''$ where the cluster density falls off.  
    	$P$-values were only calculated for a cutoff radii that include at least three points.}
        We find similar results when we relocate the cluster's centroid within the uncertainty range of \citet{Goldsbury_etal_2010}. There appears to be no central density cusp in these data.
        \label{F:cusp}}
\end{figure}

\subsection{Discussion}

The velocity dispersion profile investigated in this study does not support an IMBH detection for 47 Tuc unless one also posits a very low remnant retention fraction.
The concentrated populations of binary stars and dark stellar remnants alone appear to be enough to explain the velocity dispersion in the core.


Though the main thrust of this paper is the analysis of 47 Tuc's stellar velocity dispersion, we can also briefly look at complementary avenues of analysis. In the current state of the field, a firm case for an IMBH detection (or absence) is likely to be drawn from a suite of findings rather than any single result.

There is numerical evidence that the presence of an IMBH may cause a weak central cusp in the cluster's density/surface brightness profile \citep{Baumgardt_etal_2005,Noyola_Baumgardt_2011};
however, other works have shown that an observed cusp may also be caused by a state change in the degree of core collapse
\citep{Trenti_etal_2010,Vesperini_Trenti_2010} and is not sufficient on its own to infer an IMBH.
Our UV data show no evidence for a central cusp in the density distribution.  Figure~\ref{F:cusp} shows the cumulative distribution of objects (corrected for incompleteness) as a function of projected distance from the core. This distribution is compared against an $R^2$ relation, which would indicate a perfectly flat surface density profile.  
The mild deviation from $R^2$ in the center is very small and is not significant, as it is at the low end of the logarithmic scale and comprises very few objects
($\lesssim 10$).  
$P$-values from the Kolmogorov-Smirnov test shown in the bottom panel of Figure~\ref{F:cusp} indicate that
we cannot reject the null hypothesis (i.e. that the data are drawn from an $R^2$ distribution) to any reasonable significance until we look beyond $\sim 20''$ where the cluster density begins to fall off.

In addition, we find no sign of a small population of high-velocity stars.  If the IMBH were to become part of a binary system, three-body interactions should periodically eject high-velocity stars into the cluster on largely radial orbits.  Our data do not show any stars with velocities beyond $\sim 60$ kms$^{-1}$ while $N$-body simulations by H. Baumgardt, which are filtered to match our observational limitations, show a small population of stars with velocities up to $\sim80$ kms$^{-1}$ when they include a central IMBH.


Though we find conflicting results with the \citet{KiziltanEtal2017} study, contrary findings are nothing new in this field.  Of particular note in recent years are the contradictory
results of of \citet{Lutzgendorf_etal_2015} and \citet{Lanzoni_etal_2013} on NGC 6388,  \citet{VanderMarel_Anderson_2010} and \citet{NoyolaEtal2008} on $\omega$ Centauri, and \citet{Perera_etal_2017} and \citet{Gieles_etal_2018} on NGC 6624.
There is clearly more work to be done in characterizing and evaluating the shortcomings and strengths of different IMBH detection methods, and we could benefit from more studies that evaluate the accuracy and biases of these techniques.  For example, \citet{DeVita_etal_2017} provided insight into how using integrated-light spectroscopy might affect results versus using velocity measurements of individual objects.
Simulations can be extremely useful with their ability to explicitly incorporate a central IMBH of known mass to determine its observable effects, but the expensive computation limitations require simplified physics and assumptions about initial conditions and short-timescale dynamics. 
As has been shown in this study, some of the short-timescale phenomena such as stellar evolution/remnant retention and binary formation can play a very key role in IMBH determination.

Consideration of the sBH retention fraction led us to an unanticipated additional result of this study.  Even employing a conservative IMF to estimate the number of sBHs, we still find that a retention of $\gtrsim 8.5\%$
becomes incompatible with our velocity dispersion observations.  Above this fraction, our model fits a negative IMBH mass in an attempt to drop the velocity dispersion values in the core.
This clearly unphysical result indicates that the {model's inferred massive populations provide more concentrated mass in the core than can be explained by the the observed velocity dispersion.}
For 47 Tuc, it seems that we can place the constraint on the sBH retention to be $\lesssim 8.5\%$.


For the sake of clarity and for the guidance of future research, we would like to acknowledge a few of the areas where this study could benefit from improvement.

Including additional data sets for regions beyond the central $\sim100''$ would likely provide tighter constraints on the $M_\text{Cl}$ and $a_\text{Cl}$ parameters.  Data tight in the core are important to constrain $M_\text{IMBH}$, but the other parameters are sensitive to the entire profile beyond our field of view.  Data sets would have to be included in a careful manner to appropriately account for their differing spatial and wavelength coverages, as well as their individual completeness limitations.

Line-of-sight velocities in the core could also help to boost star counts where there are few, though these types of data have their own distinct biases and limitations to consider.

Our determination of binary counts and masses relied heavily on the use of isochrones.  Stellar evolution modeling has come a long way but is not without its uncertainties.  The effects of using different isochrones and decontaminating the binary sequence are topics that have bearing on our results but merit full studies in their own right.

There are a few aspects specific to our model that were approximated to some degree or perhaps not expanded to their fullest potential.  
{The sums of the King templates that we used for the density distributions in the Jeans equation can provide an internally consistent model (i.e.\ a distribution function $f(E)$ that satisfies Jeans' theorem) for a spherical cluster with isotropic velocity dispersion through Eddington’s formula \citep{2008gady.book.....B} for a vanishingly small BH mass if the values of $a$ do not differ by too large of a factor.
Because there is no evidence for anisotropy or rotation in the core region 
\citep{Watkins_etal_2015,Bellini_etal_2017,Heyl_etal_2017},
an isotropic distribution function $f(E)$ provides a good approximation. 
Rather than constructing $f(E)$ explicitly, we compare the values of the velocity dispersion as calculated from the model using the Jeans equation with those of the data.  
In fitting the data, we assume a normal distribution for the velocities near the center of the cluster (Eq.~\ref{E:lnL}) when fitting for the observed velocity dispersion; this final assumption is not consistent with the King template and Jeans theorem, but it does provide an excellent approximation, given the observational data.  
Furthermore, with the inclusion of the central BH, we can construct an isotropic distribution function using Eddington's formula everywhere except in the immediate vicinity of the BH (about $10''$ for the largest BH mass considered), so our isotropic models do not satisfy Jeans theorem  
there.  A more complete model could use the observed isotropy and the lack of a density cusp to provide an additional constraint on the BH mass.  
In the absence of an IMBH, we find that the concentration of the retained sBHs is sufficient to render our assumed density profile and isotropy inconsistent with Jeans theorem within about $7''$ of the center. 
For future work, a more detailed model could be constrained with data at larger radii and with the lack of anisotropy or a density cusp at small radii. }
One might also choose to set up the model to simultaneously fit the cluster's density and surface brightness profile along with the velocity dispersion for a built-in self-consistency check.

Some of the details comprising our chosen model are worth consideration as well.  For instance, we rely on the \citet{GoldsburyEtal2016} mass segregation power law to infer population concentrations for unseen objects based on their mass.  \citet{GoldsburyEtal2016} only tested this relation up to $\sim0.8 M_\odot$, so we have no guarantee that objects above this mass regime will follow the same relation. 

Additionally, one might get a better picture of the dark stellar remnant populations by doing a careful IMF evolution.  Such a process may provide more accurate masses and counts for these dark objects.

Where possible, we attempted to make conservative estimates of the concentrated mass components.  The fact that our model predicts no IMBH despite these conservative assumptions degrades any support for the idea of an IMBH in 47 Tuc according to this analysis.

We have shown here that proper motion velocity dispersion analysis can prove to be a useful tool in the search for IMBHs, even in the crowded environments of a globular cluster core.  With the right choice of photometric filters, we can overcome the crowding issues and glean a multitude of data in the region most sensitive to an IMBH's influence.   
The inclusion of concentrated populations in our mass model is sufficient to explain the observed velocity dispersion in 47 Tuc's core.
That being said, the velocity dispersion signature of an IMBH and a tight population of sBHs is very similar.  A trade-off in mass between them would not be very evident in the velocity data.  Independent estimates of the sBH retention and progenitor IMF are likely the best ways to disentangle this degeneracy.

\section*{Acknowledgements}

We would like to extend special thanks and acknowledgment to Laura Watkins, whose helpful suggestions set us on the track of looking at independent mass populations.  
We also greatly appreciate the consultation and advice of Vincent H\'{e}nault-Brunet with regards to our dynamical formulation and mass modeling.

An anonymous referee provided an enormous number of excellent suggestions, many of which were included in the final version of this paper.

The presented research used NASA/ESA \emph{Hubble
Space Telescope} observations obtained at the Space Telescope
Science Institute, which is operated by the Association of
Universities for Research in Astronomy Inc. under NASA
contract NAS5-26555. These observations are associated with
proposals GO-12971 (PI: H. Richer), GO-9443 (PI: I. King), 
and GO-10775 (PI: A. Sarajedini). This work
was supported by NASA/HST grants  GO-12971, 
the Natural Sciences and Engineering Research Council
of Canada, the Canadian Foundation for Innovation, and the
British Columbia Knowledge Development Fund. We have made
use of the NASA ADS, arXiv.org, and the Mikulski Archive
for Space Telescopes.

\clearpage


\bibliographystyle{aasjournal}
\bibliography{TucIMBH}

\end{document}